\documentclass[12pt]{article}
\usepackage{amsfonts,amssymb,epsfig}
\usepackage[latin1]{inputenc}
\usepackage[english]{babel}
\usepackage{color}
\newtheorem{thm}{Th\'eor\`eme}[section]
\newtheorem{cor}[thm]{Corollaire}
\newtheorem{lem}[thm]{Lemme}
\newtheorem{pro}[thm]{Proposition}
\newtheorem{dfn}[thm]{D\'efinition}
\newtheorem{rmk}[thm]{Remark}
\newtheorem{expl}[thm]{Exemple}

\def\dessous#1\sous#2{\mathrel{\mathop{\kern0pt#2}\limits_{#1}}}

\newcommand{\C}{\mathbb C}
\newcommand{\N}{\mathbb N}

\newcommand{\1}{1 \! \! {\rm I}}

\newcommand{\beq}{\begin{eqnarray}}
\newcommand{\eeq}{\end{eqnarray}}
\newcommand{\bpro}{\begin{pro}}
\newcommand{\epro}{\end{pro}}
\newcommand{\blem}{\begin{lem}}
\newcommand{\elem}{\end{lem}}
\newcommand{\bdfn}{\begin{dfn}}
\newcommand{\edfn}{\end{dfn}}
\newcommand{\bcor}{\begin{cor}}
\newcommand{\ecor}{\end{cor}}
\newcommand{\bthm}{\begin{thm}}
\newcommand{\ethm}{\end{thm}}
\newcommand{\bex}{\begin{expl}}
\newcommand{\eex}{\end{expl}}
\newcommand{\brmq}{\begin{rmq}}
\newcommand{\ermq}{\end{rmq}}
\newcommand{\benum}{\begin{enumerate}}
\newcommand{\eenum}{\end{enumerate}}
\newcommand{\bitem}{\begin{itemize}}
\newcommand{\eitem}{\end{itemize}}
\begin{document}

\begin{titlepage}
\begin{center}

\vspace{20pt}

{\Large\bf {Landau levels in a $2D$ noncommutative space: matrix and quaternionic vector coherent states}}


\vspace{20pt}

M. N. Hounkonnou
\footnote{Correspondence author:
norbert.hounkonnou@cipma.uac.bj  with copy to hounkonnou@yahoo.fr.} and I. Aremua

\vspace{20pt}

{\em 
International Chair of Mathematical Physics
and Applications} \\
{\em ICMPA-UNESCO Chair}\\
{\em University of Abomey-Calavi}\\
{\em  072 B.P. 50 Cotonou, Republic of Benin}\\
\vspace{10pt}

\today

\begin{abstract}

  The behavior of an electron in an external uniform electromagnetic background   coupled to a
harmonic potential, with  noncommuting space coordinates,  is considered in this work.
The  thermodynamics  of the system is studied.
Matrix vector coherent states (MVCS) as well as  quaternionic vector coherent states (QVCS), satisfying required properties,
 are also constructed and discussed.

\end{abstract}

\end{center}

\end{titlepage}

\section{Introduction}

In most of the introductory references in the literature devoted to quantum mechanics and quantum field theory,
it comes out that the natural appearance of noncommutativity in string theories has increasingly led to attempts to study physical problems in
noncommutative spaces \cite{doplicher}-\cite{douglas}. Although noncommuting coordinates are operators even at the
classical level, one can treat them as commuting by replacing
operator products by $*$-products \cite{omer-jellal1}. This approach allows one to generalize classical as well as quantum mechanics without altering
their main physical interpretations and to recover the usual results when noncommutativity is switched off.
In some recent works, the  quantum Hall system has attracted considerable attention from the point of view
of noncommutative quantum mechanics and quantum field theory (see e.g. \cite{horvathy-duval}, \cite{horvathy}, \cite{omer-jellal}, \cite{pasquier})
as it is probably the simplest physical realization of a noncommutative spatial geometry.  Note that a 
noncommutative model  valid for a constant magnetic field, with respect to the geometrical aspect of the problem,  has been also investigated. For more 
details, please see \cite{nair-polychronakos}. 
The description of such a system   \cite{goerbig, lederer}  is adequately provided by the well known
 Landau model. This latter
describes the motion of an electron in a static uniform magnetic field, studied for the
first time by Landau \cite{landau},  which can be assimilated in $2D$ to a harmonic oscillator.
Since this discovery, the quantum states of a particle in a magnetic and  electromagnetic fields   on noncommutative plane have been  attracting
considerable attention, see for instance \cite{horvathy, omer-jellal, jellal, gamboa-loewe-mendez-rojas, geloun-jan-hounkonnou, 
 dulat-li, alvarez,  ben-sunandan-scholtz}  and
 more recently \cite{zhang-horvathy} (and references listed therein). 
In \cite{jan-scholtz},
the thermodynamics  of  an ideal fermion gas in a noncommutative well in two dimensions \cite{scholtz-chakraborty-jan-vaidya},
 has been investigated. The authors have shown that  the
thermodynamical properties of the fermion gas for the commutative and noncommutative cases agree at low densities, while
 at high densities they
start diverging strongly   due to the implied excluded area resulting
from the noncommutativity.
In \cite{gazeau-hsiao-jellal} the possible occurrence of orbital magnetism for two-dimensional
electrons confined by a harmonic potential \cite{ishikawa-fukuyama} in various regimes of temperature and magnetic
field has been studied. Standard coherent states (CS)  have been used for calculating symbols of various involved
observables like the thermodynamical potential, the magnetic moment or the spatial distribution
of the current. In \cite{jellal}, an analogous treatment in a noncommutative framework has been achieved and the
results  of  \cite{gazeau-hsiao-jellal}  in the commutative case have been recovered by switching off the $\theta$-parameter.

  In the noncommutative quantum mechanics formulation, a major role is played by the CS on the quantum Hilbert space denoted by $\mathcal H_{q}$ 
(the space of
Hilbert Schmidt operators on the classical configuration space denoted by $\mathcal H_{c}$), which are expressed in terms
of a projection operator on the usual Glauber-Klauder-Sudarshan CS in the classical
configuration space. Based on  the approach developed in \cite{gouba-scholtz},   Gazeau-Klauder CS have been constructed
in the noncommutative quantum mechanics  \cite{ben-scholtz}. These
states share similar properties to those of ordinary canonical CS in the sense that they saturate the related position uncertainty relation, obey a
Poisson  distribution and possess a flat geometry.

This work deals with the Landau problem, particularly  the  study of the electron
 motion in  an external uniform electromagnetic  field coupled with  a
harmonic potential  in a two-dimensional noncommutative space.  The thermodynamics of this  physical system is investigated, not proceeding by the same way as in \cite{jan-scholtz} for  an ideal fermion gas, but
following rather  the method established
in \cite{gazeau-hsiao-jellal} by  formulating   at first  CS
 on the quantum Hilbert space $\mathcal H_{q}$.
Then, the thermodynamical potential is evaluated, by the use of deduced inequalities, together   with the   magnetic moment.
The vector coherent states (VCS) are derived; they fulfill a resolution of the identity on a suitable Hilbert space which is
consistent
with the general formulation of \cite{ali-englis-gazeau}.  We extend the VCS construction used in \cite{ben-scholtz}
to a
formal tensor product of quantum Hilbert spaces (using the primary formulation
of \cite{thirulo}), including complex matrices and quaternions as CS
variables. The physical features of the quaternionic VCS (QVCS) are discussed.

The paper is organized as follows. In Section $2$, we describe the physical model as well as
the associated matrix formulation.    The Hamiltonian spectrum and  its spectral decomposition are provided.
The definition of the passage operators from an
orthonormal basis to
another is also supplied.  Section $3$ deals with the thermodynamical aspects of the studied model.
In Section $4$,  relevant VCS  and QVCS for Landau levels are constructed
and discussed.
Finally, there follow concluding remarks
in Section $5$.
\section{The electron in noncommutative plane}
\subsection{Quantum model}

The physics of an electron  in crossed constant uniform electric  ${\bf E}$ and magnetic ${\bf B}$ fields
coupled  with a  confining harmonic potential in a noncommutative space,
 is described, in the gauge ${\bf A} = \left(-\frac{B}{2}y, \frac{B}{2}x \right),$ by the Hamiltonian:
\beq{\label{exp00}}
H_{\theta} = \frac{1}{2M}\left(\hat P_{i} - \frac{eB}{2c}\epsilon_{ij}\hat X_{j}\right)^{2}
+ \frac{M \omega^{2}_{0}}{2}\hat X^{2}_{i} - e E_{i}\hat X_{i}, \; \;
\epsilon_{ji} = -\epsilon_{ij}, \; \epsilon_{12} = +1,
\eeq
where the position and momentum operators $\hat X_{i} = \hat X, \, \hat Y$  and
$\hat P_{i} = \hat P_{X},  \, \hat P_{Y}, \, i=1,2$,  satisfy the following commutation relations of the  noncommutative Heisenberg algebra
\cite{gouba-scholtz}:
\beq
[\hat X, \hat Y] = \imath \theta, \quad [\hat X, \hat P_{X}] = \imath \hbar = [\hat Y, \hat P_{Y}] , \quad
[\hat P_{X}, \hat P_{Y}] = 0.
\eeq

The position operators $\hat X_{i}$ and their
corresponding canonically conjugate momenta $\hat P_{i}$ can be combined in the operators  $\hat \Pi_{i} = \hat P_{i}
- \frac{eB}{2c}\epsilon_{ij}\hat X_{j}$
to yield the relations:
\beq
[\hat X_{i}, \hat \Pi_{j}] = \imath  \left( \hbar- \frac{eB}{2c}\theta\right)\delta_{ij}, \quad
[\hat \Pi_{i}, \hat \Pi_{j}] = -i \frac{eB}{c}\left(\hbar  - \frac{eB}{4c}\theta\right)\epsilon_{ij}.
\eeq

From the latter, define the complex canonically conjugate momenta, denoted by $\hat \Pi_{Z}$
corresponding to $\hat Z = \hat X + \imath \hat Y$ and  $\hat{\bar Z} = \hat X - \imath \hat Y$  by
\beq
\hat \Pi_{Z} = \hat \Pi_{X} - \imath \hat \Pi_{Y}, \qquad \hat \Pi_{\bar Z} = \hat \Pi_{X} + \imath \hat \Pi_{Y},
\eeq

respectively, such that
the quantum operators $\hat Z, \hat{\bar Z} ,\hat \Pi_{Z}, \hat \Pi_{\bar Z}$ act on the quantum Hilbert space  $\mathcal H_{q}$
  \cite{ben-sunandan-scholtz, gouba-scholtz}, i.e.  the space of Hilbert-Schmidt operators
acting on the noncommutative configuration (Hilbert) space $\mathcal H_{c},$ defined as:
\beq
\mathcal H_{q} = \left\{\psi(\hat z, \hat{\bar z}): \psi(\hat z, \hat{\bar z}) \in \mathcal B(\mathcal H_{c}),\,
tr_{c}(\psi(\hat z, \hat{\bar z})^{\dag}, \psi(\hat z, \hat{\bar z}))
 < \infty \right \},
\eeq
endowed with the following inner product

\beq
(\psi(\hat x_{1}, \hat x_{2}), \phi(\hat x_{1}, \hat x_{2})) = tr_{c}(\psi(\hat x_{1}, \hat x_{2})^{\dag}, \phi(\hat x_{1}, \hat x_{2}))
\eeq

where $tr_{c}$ stands for the trace over $\mathcal H_{c}$.
$\mathcal B(\mathcal H_{c})$  is the set of bounded operators on $\mathcal H_{c}$.

\begin{rmk}
For  a harmonic oscillator, the two-dimensional noncommutative coordinate algebra is   given by
\beq
[\hat x, \hat y] = \imath \theta;
\eeq
 $\theta$ refers to as the noncommutativity parameter.  The annihilation and creation operators
$a = 1/{\sqrt{2\theta}}(\hat x + \imath \hat y),\, a^{\dag} = 1/{\sqrt{2\theta}}(\hat x - \imath \hat y)$
obey a Heisenberg-Fock
algebra $[a,a^{\dag}]  = \1_{c}$, where $\1_{c}$ is the identity operator on the Hilbert space $\mathcal H_{c},$ i.e.
the noncommutative configuration space  which becomes itself a Hilbert space isomorphic to the boson Fock space \cite{gouba-scholtz}
 $\mathcal H_{c} = span\{|n\rangle, n \in \N\}$, with $|n\rangle =  1/\sqrt{n !}(a^{\dag})^{n}|0\rangle$.
\end{rmk}

As mentioned in  \cite{ben-scholtz}, a well defined representation with self-adjoint properties with respect to 
the quantum Hilbert space $\mathcal H_{q}$ 
inner product is provided by the following relations

\beq
\hat X \psi = \hat x \psi, \quad \hat Y \psi = \hat y \psi, \quad \hat P_{X}\psi = \frac{\hbar}{\theta}[\hat y,  \psi],  \quad 
\hat P_{Y}\psi = -\frac{\hbar}{\theta}[\hat x,  \psi].
\eeq

  On the Hilbert space  $\mathcal H_{q}$,   the following commutation relations are satisfied:

\beq{\label{delt00}}
[\hat Z, \hat{\bar Z}] &=& 2\theta, \qquad  [\hat Z, \hat \Pi_{Z}] = 2\imath\left( \hbar - \frac{eB}{2 c}\theta\right)
= [\hat{\bar Z}, \hat \Pi_{\bar Z}], \crcr
 [\hat \Pi_{Z}, \hat \Pi_{\bar Z}] &=& 2 \frac{eB}{c} \left(\hbar -\frac{eB}{4c}\theta\right).
\eeq

For the analysis purpose, defining   a diagonal matrix $\mathcal D $ and  adopting the notations $E = (E_{1}, E_{2},0,0),
  {\mathcal X_{0}} = ( x_{0},  y_{0},0,0)^{t}$,
where  $ x_{0} = \frac{eE_{1}}{M \omega^{2}_{0}}, \; y_{0} = \frac{eE_{2}}{M \omega^{2}_{0}}$, the Hamiltonian $H_{\theta}$ of
the physical model can be rewritten in a short form as follows:
\beq
H_{q} = \frac{1}{4M} \hat{\mathcal Z}^{\ddag}\hat{\mathcal Z} - \frac{1}{2} e E.{\mathcal X_{0}}
= \frac{1}{4M} A^{\ddag}\mathcal D A - \frac{1}{2} e  E. {\mathcal X_{0}}, \quad A = (B_{+}, B^{\ddag}_{+}, B_{-}, B^{\ddag}_{-}).
\eeq
 The symbol $t$ means the transpose operation.

 Now introduce the operators

\beq
A^{+} &=& (B^{\ddag}_{+}, B_{+}, B^{\ddag}_{-}, B_{-})^{t} = \Lambda A, \crcr
\hat \mathcal Z^{+} &=& (\hat{\bar Z'} - {\bar  Z'_{0}}, \hat Z' -  Z'_{0}, \hat \Pi_{\bar Z} -  \Pi_{\bar Z_{0}},
\hat \Pi_{Z} -  \Pi_{Z_{0}})^{t} = \Lambda \hat{\mathcal Z}
\eeq

where $\hat Z' -  Z'_{0} = M\omega_{0} (\hat Z -  Z_{0})$,  with the permutation matrix $\Lambda$   defined by
\beq
\Lambda = \left(\begin{array}{cccc}
0 & 1 & 0 & 0 \\
1 & 0 & 0 & 0 \\
0 & 0 & 0 & 1 \\
0 & 0 & 1 & 0
                \end{array}
\right)
\eeq
and reserving the notation $\ddag$   to denote the  Hermitian conjugation on the  quantum Hilbert space.

Then, consider the  matrix $\mathfrak g$ with entries
${\mathfrak g}_{lk} = [\hat{\mathcal  Z}_{l}, \hat{\mathcal Z}^{+}_{k}], \; l, k = 1,\dots, 4,$  obtained from the commutation relations
 (\ref{delt00}) as follows:

\beq
\mathfrak g =
&& \left(\begin{array}{cccc}
2M^{2}\omega^{2}_{0}\theta & 0 & 0 & 2\imath \hbar M\omega_{0}\times \\
  &     &    &  \times \left(1-\frac{M\omega_{c}}{2\hbar}\theta\right)\\
0  & -2M^{2}\omega^{2}_{0}\theta  & 2\imath \hbar M\omega_{0} \times  & 0 \\
&   &    \times \left(1-\frac{M\omega_{c}}{2\hbar}\theta\right)  &\\
0 & -2\imath \hbar M\omega_{0} \times  & 2\hbar M\omega_{c} \times
 & 0  \\
& \times \left(1-\frac{M\omega_{c}}{2\hbar}\theta\right) & \times \left(1-\frac{M\omega_{c}}{4\hbar}\theta\right) \\
-2\imath \hbar M\omega_{0} \times   & 0 & 0 & -2\hbar M\omega_{c} \times
 \\
\times \left(1-\frac{M\omega_{c}}{2\hbar}\theta\right) & & &  \times \left(1-\frac{M\omega_{c}}{4\hbar}\theta\right)\\
\end{array}
\right) \nonumber \\
\eeq
with the eigenvalues $\tilde \lambda_{\pm}, -\tilde \lambda_{\pm}$  supplied
by the expressions

\beq
\tilde \lambda_{\pm} &=& M\hbar \left\{\Omega \sqrt{1-\frac{M\omega_{c}}{2\hbar}\theta + \left(\frac{M \Omega}{2\hbar}\theta\right)^{2}} \pm
 \omega_{c}\left(1 - \left(\frac{\omega_{c}}{4\hbar} + \frac{\omega^{2}_{0}}{\hbar \omega_{c}}\right)M\theta\right)\right \}
\eeq

where $\Omega^{2} = 4\omega^{2}_{0} +  \omega^{2}_{c}$. The  matrix $\mathcal S^{\dag}$,   eigenvector matrix of $\mathfrak g$, is given by

\beq
\mathcal S^{\dag} = \left(\frac{1}{\sqrt{|\lambda_{+}|}}u'_{1}, \frac{1}{\sqrt{|\lambda_{+}|}}(\Lambda u^{*}_{1})',
\frac{1}{\sqrt{|\lambda_{-}|}}u'_{2},  \frac{1}{\sqrt{|\lambda_{-}|}}(\Lambda u^{*}_{2})'\right)
\eeq

where the  normalized eigenvectors $(u'_{1}, (\Lambda u^{*}_{1})')$ and $(u'_{2}, (\Lambda u^{*}_{2})')$ associated with
$(\tilde \lambda_{+}, -\tilde \lambda_{+})$ and $( \tilde \lambda_{-}, - \tilde \lambda_{-})$, respectively,   are given by

\beq
u'_{1} = \frac{1}{||u_{1}||}u_{1}, \; (\Lambda u^{*}_{1})'  = \frac{1}{||\Lambda u^{*}_{1}||}[\Lambda u^{*}_{1}], \;
u'_{2} = \frac{1}{||u_{2}||}u_{2}, \; (\Lambda u^{*}_{2})'  = \frac{1}{||\Lambda u^{*}_{2}||}[\Lambda u^{*}_{2}]
\eeq

with

\beq
u_{1} = \left(\begin{array}{c}
0 \\
\imath \frac{B_{\hbar}}{\kappa_{+}} \\
1 \\
0
              \end{array}
\right), \qquad
u_{2} = \left(\begin{array}{c}
\imath \frac{B_{\hbar}}{\kappa_{-}} \\
0 \\
0 \\
1
              \end{array}
\right);
\eeq

$u^{*}_{j}, j = 1,2,$  are  the vectors with entries which are conjugate of the   $u_{j}$ entries;
$ B_{\hbar} = 2 \hbar M\omega_{0}\left(1-\frac{M\omega_{c}}{2\hbar}\theta\right)$ and

\beq
 \kappa_{\pm} = M\hbar \left\{\Omega \sqrt{1-\frac{M\omega_{c}}{2\hbar}\theta + \left(\frac{M \Omega}{4\hbar}\theta\right)^{2}} \pm
 \omega_{c}\left(1 - \left(\frac{\omega_{c}}{4\hbar} - \frac{\omega^{2}_{0}}{\hbar \omega_{c}}\right)M\theta\right)\right \}.
\eeq

Then, the Hamiltonian is obtained as

\beq
H_{q} = \frac{1}{4M} A^{\ddag}\mathcal D A - \frac{1}{2} e  E. {\mathcal X_{0}} =
\frac{1}{4M} A^{\ddag} \mathbb J_{4}\mathcal S\mathfrak g^{2}\mathcal S^{\dag}\mathbb J_{4}
A - \frac{1}{2} e  E. {\mathcal X_{0}}
\eeq

where $\mathbb J_{4}$  is given by
\beq
\mathbb J_{4}  = \mbox{diag}(\sigma_{3}, \sigma_{3}), \qquad \sigma_{3} = \left(
\begin{array}{cc}
1 & 0 \\
0 & -1 \\
\end{array}
\right) .
\eeq

Setting $\tilde \Omega_{\pm} = \frac{\tilde \Omega \pm \tilde \omega_{c}}{2}$,  where the $\theta$-dependent quantities
$\tilde \Omega$ and $\tilde \omega_{c}$
are given by
\beq
\tilde \Omega = \Omega \sqrt{1-\frac{M\omega_{c}}{2}\theta + \left(\frac{M \Omega}{4}\theta\right)^{2}}, \quad
\tilde \omega_{c} = \omega_{c}\left(1 - \left(\frac{\omega_{c}}{4} + \frac{\omega^{2}_{0}}{ \omega_{c}}\right)M\theta\right),
\eeq
then we can re-express the Hamiltonian $H_q$ in terms of positive quantities $\tilde \Omega_\pm$ as
follows:
\beq
H_{q} = \frac{\hbar}{2} \left(\tilde \Omega_{+}\tilde N_{+} + \tilde \Omega_{-}\tilde N_{-} +  \tilde \Omega \right)
 - \frac{1}{2}e(E_{1}x_{0} +
E_{2}y_{0}),
\eeq

where  $\tilde N_{\pm} = B^{\ddag}_{\pm}B_{\pm}$ denote the number operators on the quantum Hilbert space;
$B^{\ddag}_{\pm}, B_{\pm}$ are the corresponding creation and annihilation operators.

Further defining the quantities
\beq
\zeta =  \sqrt{\frac{M \Omega}{\hbar}} \frac{1}{\mu_{\theta}}
= \sqrt[4]{\frac{(M \Omega/\hbar)^{2}}{1-\frac{M\omega_{c}}{2}\theta + \left(\frac{M \Omega}{4}\theta\right)^{2}}}, \qquad
\mu_{\theta} =  \sqrt[4]{1 - \frac{M\omega_{c}}{2}\theta + \left(\frac{M\Omega}{4} \theta\right)^{2}}
\eeq

which are also $\theta-$ dependent functions, the  annihilation and creation  operators are deduced as
\beq
B_{+} &=& \zeta \frac{\hat{\bar{Z}} - {\bar{Z}}_{0}}{2} + \frac{\imath}{\zeta \hbar}(\hat P_{Z} - P_{Z_{0}}), \qquad
B^{\ddag}_{+} = \zeta \frac{\hat Z - Z_{0}}{2} - \frac{\imath}{\zeta \hbar}(\hat P_{\bar Z} -  P_{\bar Z_{0}}) \crcr
B_{-} &=&  \zeta \frac{\hat Z -  Z_{0}}{2} +
\frac{\imath}{\zeta \hbar}(\hat P_{\bar Z} -  P_{\bar Z_{0}}), \qquad  B^{\ddag}_{-} =  \zeta \frac{\hat{\bar{Z}} -
{\bar{Z}}_{0}}{2} -
\frac{\imath}{\zeta \hbar}(\hat P_{Z} -  P_{Z_{0}})
\eeq
satisfying the commutation relations:
\beq{\label{commtation}}
[B_{\pm}, B^{\ddag}_{\pm}] = \1_{q}, \;\;\; [B_{\pm}, B^{\ddag}_{\mp}] = 0, \;\;\; [B_{+}, B_{-}] = 0,
\;\;\; [B^{\ddag}_{+}, B^{\ddag}_{-}] = 0.
\eeq

Finally, there result
the eigenvalues of the Hamiltonian $H_{q},$  expressed in the Fock helicity  representation  $|\tilde n_{+}, \tilde n_{-}\rangle $   by
\beq
E_{\tilde n_{+}, \tilde n_{-}} &=& \frac{\hbar}{2}
\left(\tilde \Omega_{+}\tilde n_{+} + \tilde \Omega_{-}\tilde n_{-} +  \tilde \Omega \right)
 - \frac{1}{2}e(E_{1}x_{0} +
E_{2}y_{0})
\eeq
with the corresponding eigenvectors
on the quantum Hilbert space given by
\beq
|\tilde n_{+}, \tilde n_{-})
= \frac{1}{\sqrt{\tilde n_{+} !\tilde n_{-} !}}\left(B^{\ddag}_{+}\right)^{\tilde n_{+}}
\left(B^{\ddag}_{-}\right)^{\tilde n_{-}}|0\rangle \langle 0|
,
\eeq
where $B^{\ddag}_{-}$ may have an action on the right by $B_{-}$ on  $|0\rangle \langle 0|$ which
stands for the vacuum state on $\mathcal H_{q}$ and $|||\tilde n_{+}, \tilde n_{-})|| = 1$.

The  annihilation and creation  operators act  on the states $|\tilde n_{+}, \tilde n_{-}) =
|\tilde n_{+} \rangle \langle \tilde n_{-}| , \;  \tilde n_{\pm} = 0, 1, 2, \dots,$
as follows:

\beq
B_{+}|\tilde n_{+}, \tilde n_{-}) &=&  \sqrt{\tilde n_{+}}|\tilde n_{+}-1, \tilde n_{-}),
\qquad  B^{\ddag}_{+}|\tilde n_{+}, \tilde n_{-}) = \sqrt{\tilde n_{+} + 1 }|\tilde n_{+}+1, \tilde n_{-}), 
\eeq
\beq
B_{-}|\tilde n_{+}, \tilde n_{-}) &=&  
\sqrt{\tilde n_{-}}|\tilde n_{+}, \tilde n_{-}-1),
\qquad  B^{\ddag}_{-}|\tilde n_{+}, \tilde n_{-}) =  \sqrt{\tilde n_{-} + 1 } |\tilde n_{+}, \tilde n_{-}+1).
\eeq

\subsection{Spectral decomposition}

Let us consider the  dimensionless shifted quantum Hamiltonian
\beq
H^{dim}_{q} = \frac{1}{\hbar \tilde \Omega}
\left[H_{q} +  \frac{1}{2}e(E_{1}x_{0} +
E_{2}y_{0}) \right]
\eeq
with associated eigenvalues

\beq
\tilde E_{\tilde n_{+}, \tilde n_{-}}
=  \frac{1}{2}\left(\frac{\tilde \Omega_{+}}{\tilde \Omega}\tilde n_{+} + \frac{\tilde \Omega_{-}}{\tilde \Omega}\tilde n_{-} + 1\right) .
\eeq

Take $\{|\tilde n_{+}, \tilde n_{-}), \tilde n_{\pm} \in \N\}$
as  the orthonormal eigenstate basis associated with
the quantum Hamiltonian  $ H_{q}$ in the helicity Fock algebra representation. With respect to the inner product on $\mathcal H_{q}$,
we have
$(\tilde n_{+}, \tilde n_{-}|\tilde n'_{+}, \tilde n'_{-})
= {\mbox tr}_{c}[(|\tilde n_{+} \rangle \langle \tilde n_{-}|)^{\ddag}|\tilde n'_{+} \rangle \langle\tilde n'_{-}|]
= \delta_{\tilde n_{+},\tilde n'_{+}}
\delta_{\tilde n_{-},\tilde n'_{-}}$.
On this basis, the Hamiltonian $H^{dim}_{q}$ admits the following
spectral decomposition

\beq
H^{dim}_{q} = \sum_{\tilde n_{\pm}=0}^{\infty}
|\tilde n_{+}, \tilde n_{-}) \tilde E_{\tilde n_{+}, \tilde n_{-}}(\tilde n_{+}, \tilde n_{-}|.
\eeq

Let $\{|n\rangle \langle m| := |n,m), n, m \in \N \}$
be  the orthonormal basis associated with  the quantum Hilbert space $\mathcal H_{q}$. Introduce the passage operators from
$\{|n,m), n, m \in \N \}$ to $\{|\tilde n_{+}, \tilde n_{-}), \tilde n_{\pm} \in \N \}$ and vice versa given by

\beq
\mathcal U|n,m) = |\tilde n_{+}, \tilde n_{-}), \qquad \mathcal V|\tilde n_{+}, \tilde n_{-}) = |n,m)
\eeq

where their expansions are given by

\beq
\mathcal U = \sum_{n,m=0}^{\infty} |\tilde n_{+}, \tilde n_{-})(n,m|, \qquad
\mathcal V = \sum_{\tilde n_{\pm}=0}^{\infty} |n,m)(\tilde n_{+}, \tilde n_{-}|,
\eeq

respectively. $\mathcal U, \mathcal V$ are obtained as mutually adjoint through  the following identities satisfied
  on $\mathcal H_{q}:$

\beq
\mathcal U \mathcal V = \sum_{\tilde n_{\pm}=0}^{\infty}|\tilde n_{+}, \tilde n_{-})(\tilde n_{+}, \tilde n_{-}|
=  \mathbb I_{q},
\qquad \mathcal V \mathcal U = \sum_{n,m=0}^{\infty}|n,m)(n,m| =  \mathbb I_{q},
\eeq

where $\mathbb I_{q}$ stands for the  identity  on $\mathcal H_{q}$.
Then, the Hamiltonian $H^{dim}_{q}$ can be rewritten in a diagonal form as below:

\beq{\label{hq00}}
\mathbb H^{dim}  = \mathcal V H^{dim}_{q} \mathcal U =  \sum_{n,m=0}^{\infty}|n,m)\tilde E_{n,m}(n,m|, \qquad
\tilde E_{n,m} = \frac{1}{2}\left(\frac{\tilde \Omega_{+}}{\tilde \Omega}n + \frac{\tilde \Omega_{-}}{\tilde \Omega}m + 1 \right).
\eeq

\section{Coherent states and thermodynamics of the model}

 For the   Hamiltonian
$H_{q}$  with eigenvalues
$E_{\tilde{n}_{+}, \tilde{n}_{-}}=
\frac{\hbar}{2}
\left(\tilde \Omega_{+}\tilde n_{+} + \tilde \Omega_{-}\tilde n_{-} +  \tilde \Omega \right)  + k_{e,E},
\, k_{e,E}= - \frac{1}{2}e\left(E_{1}x_{0}  +
E_{2}y_{0} \right), $ the
 coherent states denoted by $|z_{\pm},\tau)$ are defined
on the quantum Hilbert space  $\mathcal H_{q}$,  as follows:
\beq{\label{vect00}}
|z_{\pm},\tau) &=& \mathbb U(\tau)|z_{+}\rangle  \langle z_{-}| \cr
 &=&   e^{-\frac{1}{2}(|z_{+}|^{2} + |z_{-}|^{2})}
\sum_{\tilde{n}_{+}, \tilde{n}_{-} = 0}^{\infty}
\frac{z_{+}^{\tilde{n}_{+}}\bar z_{-}^{\tilde{n}_{-}}}{\sqrt{\tilde{n}_{+} !\tilde{n}_{-} !}}
e^{-i \tau  E_{\tilde n_{+}, \tilde n_{-}}}
|\tilde{n}_{+}\rangle \langle \tilde{n}_{-}|.
\eeq

The parameter $\tau$ is introduced such that the states (\ref{vect00})  fulfill the Gazeau-Klauder axiom
of temporal stability  relative to the classical time evolution operator
$\mathbb U(\tau) = e^{-\imath \left[H_{q}\right]\tau }$. Indeed, we have the following.

\bpro
These vectors satisfy the following properties:
\bitem
\item [-] Temporal stability

\beq
\mathbb U(t)|z_{\pm}, \tau)  =   e^{-\imath \left[H_{q}\right]t }|z_{\pm}, \tau)
= |z_{\pm}, \tau + t) ,
\eeq

\item [-] Action identity, also called lower symbol of
$H_{q}$,
\beq{\label{ncident}}
\mbox{\v{H}}_{q}(z_{\pm}) = (z_{\pm}, \tau|H_{q}|z_{\pm},  \tau)
= \frac{\hbar}{2} \left(\tilde \Omega_{+}|z_{+}|^{2} + \tilde \Omega_{-}|z_{-}|^{2} + \tilde \Omega\right) +   k_{e,E},
\eeq

\item [-] Resolution of the identity
\beq{\label{resolv}}
\frac{1}{\pi^{2}}\int_{\C^{2}}|z_{\pm},\tau)
(z_{\pm}, \tau|d^{2}z_{+}d^{2}z_{-} \equiv \mathbb I_{q},
\eeq
where $\mathbb I_{q}$ is the identity operator on $ \mathcal H_{q}$ provided by

\beq{\label{resolv01}}
\mathbb I_{q} = \frac{1}{\pi}\int_{\C}dzd\bar{z}|z)e^{\overleftarrow{\partial_{\bar z}}\overrightarrow{\partial_{z}}} (z|.
\eeq

\eitem
\epro

{\bf Proof:}

 We have from the  definition (\ref{vect00}) the following relation 
\beq
\frac{1}{\pi^{2}}\int_{\C^{2}}d^{2}z_{+}d^{2}z_{-}|z_{+}\rangle \langle z_{+}|  |z_{-}\rangle \langle z_{-}|  \equiv 
\frac{1}{\pi^{2}}\int_{\C^{2}}d^{2}z_{+}d^{2}z_{-}|z_{\pm}) (z_{\pm}| 
\eeq

where from the definition (\ref{vect00}), we have

\beq
|z_{\pm} ) &=&  |z_{+}\rangle  \langle z_{-}| \cr
 &=&   e^{-\frac{1}{2}(|z_{+}|^{2} + |z_{-}|^{2})}
\sum_{\tilde{n}_{+}, \tilde{n}_{-} = 0}^{\infty}
\frac{z_{+}^{\tilde{n}_{+}}\bar z_{-}^{\tilde{n}_{-}}}{\sqrt{\tilde{n}_{+} !\tilde{n}_{-} !}} |\tilde{n}_{+}\rangle \langle \tilde{n}_{-}|.
\eeq



By taking a state $|\psi)$ on  $ \mathcal H_{q}$, we obtain

\beq
\frac{1}{\pi^{2}}\int_{\C^{2}}d^{2}z_{+}d^{2}z_{-}|z_{\pm}) (z_{\pm}|\psi) &=& \frac{1}{\pi^{2}}\int_{\C^{2}}d^{2}z_{+}d^{2}z_{-}
|z_{+}\rangle \langle z_{-}| \sum_{\tilde n_{\pm}=0}^{\infty}|\tilde n_{-}\rangle\langle \tilde n_{+}
|[|z_{+} \rangle \langle z_{-}|]^{\ddag} \psi|\tilde n_{+} \rangle \langle \tilde n_{-}| \cr
&=& \frac{1}{\pi^{2}}\int_{\C^{2}}d^{2}z_{+}d^{2}z_{-}
|z_{+}\rangle \langle z_{-}| \langle z_{+}|\psi|z_{-} \rangle \cr
&=& |\psi)
\eeq

such that

\beq{\label{ncres01}}
\frac{1}{\pi^{2}}\int_{\C^{2}}d^{2}z_{+}d^{2}z_{-}|z_{\pm}) (z_{\pm}| \equiv \mathbb I_{q}.
\eeq

In order to provide an equivalence between (\ref{resolv}) and (\ref{resolv01}), let us  consider the following relations

\beq{\label{resolv02}}
\mathbb I_{q}|\psi) &=& \frac{1}{\pi^{2}}\int_{\C^{2}}dzd\bar{z}dwd\bar{w}
|z\rangle \langle w| \langle z|\psi|w \rangle \cr
&=& \frac{1}{\pi^{2}}\int_{\C^{2}}dzd\bar{z}dud\bar{u}
|z\rangle \langle z+u| \langle z|\psi|z+u \rangle \cr
&=& \frac{1}{\pi}\int_{\C}dzd\bar{z}\frac{1}{\pi}\int_{\C}d^{2}ue^{-|u|^{2}}|z\rangle\langle z|
e^{\bar u\overleftarrow{\partial_{\bar z}} + u\overrightarrow{\partial_{z}}}\langle z|\psi|z\rangle, 
\eeq

where   $w = z + u$ with $d^{2}w = d^{2}u $,  and   $e^{u \partial_{z}} f(z) = f(z+u)$.  Then,  set

\beq
\frac{1}{\pi}\int_{\C}d^{2}ue^{-|u|^{2}}|z\rangle\langle z|
e^{\bar u\overleftarrow{\partial_{\bar z}} + u\overrightarrow{\partial_{z}}}\langle z|\psi|z\rangle =
\frac{1}{\pi}\int_{\C}d^{2}ue^{-|u|^{2}}|z\rangle\langle z|e^{\bar u\overleftarrow{\partial_{\bar z}}}e^{u \overrightarrow{\partial_{z}}}
\langle z|\psi|z\rangle
\eeq

and

\beq
I = |z\rangle\langle z|e^{\bar u\overleftarrow{\partial_{\bar z}}}e^{u \overrightarrow{\partial_{z}}}
\langle z|\psi|z\rangle.
\eeq

We have

\beq
I &=& \left[\sum_{n', m' = 0}^{\infty}|n'\rangle\langle m'|e^{-{\bar z}z}\frac{{\bar z}^{m'}}{\sqrt{m' !}}\frac{{ z}^{n'}}{\sqrt{n' !}}\right]
e^{\bar u\overleftarrow{\partial_{\bar z}}}e^{u \overrightarrow{\partial_{z}}}
\left[\sum_{n, m = 0}^{\infty}|n\rangle\langle m|e^{-{\bar z}z}\frac{{\bar z}^{n}}{\sqrt{n !}}\frac{{ z}^{m}}{\sqrt{m !}}\right] \crcr
&=& \left[ \sum_{n', m' = 0}^{\infty}\frac{z^{n'}}{\sqrt{n' !}}\frac{{\bar z}^{n}}{\sqrt{n !}}|n'\rangle\langle m'| \right]
\left(e^{-{\bar z}z} \frac{{\bar z}^{m'}}{\sqrt{m' !}}\right)
e^{\bar u\overleftarrow{\partial_{\bar z}}}e^{u \overrightarrow{\partial_{z}}}\left(e^{-{\bar z}z} \frac{{z}^{m}}{\sqrt{m !}}\right).
\eeq

Let 

\beq
K(z) = \left(e^{-{\bar z}z} \frac{{\bar z}^{m'}}{\sqrt{m' !}}\right)
e^{\bar u\overleftarrow{\partial_{\bar z}}}e^{u \overrightarrow{\partial_{z}}}\left(e^{-{\bar z}z} \frac{{z}^{m}}{\sqrt{m !}}\right).
\eeq

We obtain

\beq
K(z) &=& \frac{1}{\sqrt{m' !}} \frac{1}{\sqrt{m !}}\sum_{k=0}^{\infty}\sum_{l=0}^{\infty}\frac{1}{k !}
\left({\bar u}^{k}\partial^{k}_{{\bar z}}[{\bar z}^{m'}
e^{-{\bar z}z}]\right)\frac{1}{l !}\left({u}^{l}\partial^{l}_{{z}}\left[{ z}^{m}
e^{-{\bar z}z}\right]\right)
\eeq

which supplies, by performing a radial parametrization,  that 

\beq
&&\frac{1}{\pi}\int_{\C}d^{2}ue^{-|u|^{2}}K(z) \cr
&=& \frac{1}{\sqrt{m' !}} \frac{1}{\sqrt{m !}}\sum_{k=0}^{\infty}\sum_{l=0}^{\infty}
\frac{1}{\pi}\int_{\C}d^{2}ue^{-|u|^{2}}\frac{{\bar u}^{k}}{k !}\frac{u^{l}}{l !}\partial^{k}_{{\bar z}}[{\bar z}^{m'}
e^{-{\bar z}z}]\partial^{l}_{{z}}\left[{ z}^{m}
e^{-{\bar z}z}\right] \cr
&=&  \frac{1}{\sqrt{m' !}} \frac{1}{\sqrt{m !}}\sum_{k=0}^{\infty}\sum_{l=0}^{\infty}
\frac{1}{\pi}\int_{0}^{\infty}rdre^{-r^{2}}\frac{r^{k+l}}{k ! l !}\int_{0}^{2\pi}e^{-\imath(l-k)\phi}d\phi \cr
&& \times \partial^{k}_{{\bar z}}[{\bar z}^{m'}
e^{-{\bar z}z}]\partial^{l}_{{z}}\left[{ z}^{m}
e^{-{\bar z}z}\right]\cr
&=&   \frac{1}{\sqrt{m' !}} \frac{1}{\sqrt{m !}}\sum_{k=0}^{\infty}\left[\frac{1}{k !}\int_{0}^{\infty}2r^{2k+1}e^{-r^{2}}dr\right]
\left[\frac{1}{k !}\partial^{k}_{{\bar z}}[{\bar z}^{m'}
e^{-{\bar z}z}]\partial^{k}_{{z}}\left[{ z}^{m}
e^{-{\bar z}z}\right]\right] \cr
&=& \frac{1}{\sqrt{m' !}} \frac{1}{\sqrt{m !}}\sum_{k=0}^{\infty}\left[\frac{1}{k !}\partial^{k}_{{\bar z}}[{\bar z}^{m'}
e^{-{\bar z}z}]\partial^{k}_{{z}}\left[{ z}^{m}
e^{-{\bar z}z}\right]\right]. 
\eeq

Besides, 

\beq
 \left(e^{-{\bar z}z} \frac{{\bar z}^{m'}}{\sqrt{m' !}}\right)
e^{\overleftarrow{\partial_{\bar z}}\overrightarrow{\partial_{z}}}\left(e^{-{\bar z}z} \frac{{z}^{m}}{\sqrt{m !}}\right) = 
\frac{1}{\sqrt{m' !}} \frac{1}{\sqrt{m !}}\sum_{k=0}^{\infty}\left[\frac{1}{k !}\partial^{k}_{{\bar z}}[{\bar z}^{m'}
e^{-{\bar z}z}]\partial^{k}_{{z}}\left[{ z}^{m}
e^{-{\bar z}z}\right]\right]
\eeq

which implies that 

\beq
\frac{1}{\pi}\int_{\C}d^{2}ue^{-|u|^{2}}K(z) =  \left(e^{-{\bar z}z} \frac{{\bar z}^{m'}}{\sqrt{m' !}}\right)
e^{\overleftarrow{\partial_{\bar z}}\overrightarrow{\partial_{z}}}\left(e^{-{\bar z}z} \frac{{z}^{m}}{\sqrt{m !}}\right).
\eeq

Then, 

\beq
&&\frac{1}{\pi}\int_{\C}d^{2}ue^{-|u|^{2}}|z\rangle\langle z|
e^{\bar u\overleftarrow{\partial_{\bar z}} + u\overrightarrow{\partial_{z}}}\langle z|\psi|z\rangle \cr
&=& 
\left[ \sum_{n', m' = 0}^{\infty}\frac{z^{n'}}{\sqrt{n' !}}\frac{{\bar z}^{n}}{\sqrt{n !}}|n'\rangle\langle m'| \right]
\left(e^{-{\bar z}z} \frac{{\bar z}^{m'}}{\sqrt{m' !}}\right)
e^{\overleftarrow{\partial_{\bar z}}\overrightarrow{\partial_{z}}}\left(e^{-{\bar z}z} \frac{{z}^{m}}{\sqrt{m !}}\right)\cr
&=& \left[\sum_{n', m' = 0}^{\infty}|n'\rangle\langle m'|e^{-{\bar z}z}\frac{{\bar z}^{m'}}{\sqrt{m' !}}\frac{{ z}^{n'}}{\sqrt{n' !}}\right]
e^{\overleftarrow{\partial_{\bar z}}\overrightarrow{\partial_{z}}}
\left[\sum_{n, m = 0}^{\infty}|n\rangle\langle m|e^{-{\bar z}z}\frac{{\bar z}^{n}}{\sqrt{n !}}\frac{{ z}^{m}}{\sqrt{m !}}\right] \cr
&=& |z)e^{\overleftarrow{\partial_{\bar z}}\overrightarrow{\partial_{z}}} (z|.
\eeq

Thus, (\ref{resolv02}) becomes for a given operator $|\psi)$

\beq
\mathbb I_{q}|\psi) &=& \frac{1}{\pi}\int_{\C}dzd\bar{z}\frac{1}{\pi}\int_{\C}d^{2}ue^{-|u|^{2}}|z\rangle\langle z|
e^{\bar u\overleftarrow{\partial_{\bar z}} + u\overrightarrow{\partial_{z}}}\langle z|\psi|z\rangle \cr
&=& \frac{1}{\pi}\int_{\C}dzd\bar{z}|z)e^{\overleftarrow{\partial_{\bar z}}\overrightarrow{\partial_{z}}} (z|\psi)
\eeq

which completes the proof.

$\hfill{\square}$

Provided the definition of the  upper (or covariant)
symbol \cite{gazeau-hsiao-jellal, jellal} of  an appropriate observable $\mathcal O$,   given by

\beq
\mathcal O = \frac{1}{\pi^{2}}\int_{\C^{2}} \hat{\mathcal O}|z_{\pm},\tau)
(z_{\pm}, \tau| d^{2}z_{+}d^{2}z_{-},
\eeq
the upper symbol of the Hamiltonian $H_{q}$ is furnished by the formula

\beq{\label{ncident0}}
\mbox{\^{H}}_{q}(z_{\pm})
= \frac{\hbar}{2} \left(\tilde \Omega_{+}|z_{+}|^{2} + \tilde \Omega_{-}|z_{-}|^{2} - \tilde \Omega\right) + k_{e,E}.
\eeq

Assume that the term $ k_{e,E} = - \frac{1}{2}e(E_{1} x_{0}  +
E_{2} y_{0})$ is a mere constant, and set $H_{q} = \mathcal H_{OSC} +
\frac{\tilde \omega_{c}}{2}L_{z} +  k_{e,E}$, such that $\mathcal H_{OSC}$ and $L_{z}$ are given by
\beq{\label{exp07}}
\mathcal H_{OSC} &=& \frac{1}{2M}(\tilde p_{x} - \tilde p_{x_{0}})^{2} + \frac{1}{2M}(\tilde p_{y} - \tilde p_{y_{0}})^{2}
+ \frac{M\Omega^{2}}{8}\left[\left(x  - x_{0} \right)^{2}
+ \left(y  - y_{0} \right)^{2} \right], \crcr
L_{z} &=&  (x  - x_{0})(p_{y} - p_{y_{0}}) - (y  - y_{0})(p_{x}-p_{x_{0}}),
\eeq
with  $p_{x_{0}} = -\frac{eB}{2}y_{0}, \quad p_{y_{0}} = \frac{eB}{2}x_{0}$ and
$\tilde p^{2}_{\nu} = \left(1 - \frac{M\omega_{c}}{2}\theta
+ \left(\frac{M\Omega}{4} \theta\right)^{2}\right) p^{2}_{\nu}, \quad \nu = x, y.$ Then,  setting $\Psi(r,\varphi) = R(r)e^{\imath \rho \varphi}$,
where the  polar coordinates  $(x,y) = (r \sin{\varphi}, r\cos{\varphi})$ with $0<r<\infty $ and $0 \leq \varphi \leq \pi$ are introduced,
the stationary  Schr\"{o}dinger equation
$\mathcal H \Psi = \mathcal E \Psi$, where $\mathcal H = \mathcal H_{OSC} +
\frac{\tilde \omega_{c}}{2}L_{z}$,
is detailed as follows:
\beq
&&\left[-\frac{\hbar^{2}}{2M}\left(1-\frac{M\omega_{c}}{2}\theta + \left(\frac{M \Omega}{4}\theta\right)^{2}\right)
\left(\partial^{2}_{r} + \frac{1}{r}\partial_{r} +
\frac{1}{r^{2}}\partial^{2}_{\varphi}\right) -\imath \frac{\hbar}{2}\tilde \omega_{c}\partial_{\varphi}
+ \frac{M\Omega^{2}}{8}r^{2}  \right]\Psi(r,\varphi) \cr
&&= \mathcal E\Psi(r,\varphi),
\eeq
providing
the eigenstates and eigenvalues

\beq{\label{eigens00}}
\Psi_{n,\rho}(r,\varphi) = (-1)^{n} \sqrt{\frac{\xi}{\pi}} \sqrt{\frac{n !}{(n+|\rho|) !}}
\exp\left\{-\frac{\xi r^{2}}{2}\right\} \left(\sqrt{\xi}r\right)^{|\rho|} L^{(|\rho|)}_{n,\theta}(\xi r^{2}) e^{\imath \rho \varphi},
\eeq
and
\beq{\label{eigens01}}
\mathcal E_{n,\rho, \theta} = \hbar \tilde \Omega \left(n + \frac{|\rho| + 1}{2}\right) + \frac{\hbar \tilde \omega_{c}}{2} \rho
 - \frac{1}{2} e(E_{1}x_{0} + E_{2}y_{0}),
\eeq
respectively. Here
the
\beq
L^{(|\rho|)}_{n,\theta}(\xi r^{2}) = \sum_{m=0}^{n} (-1)^{m} \left(\begin{array}{c}
                                                                      n + |\rho| \\
                                                                      n-m
                                                                     \end{array}
\right) \frac{(\xi r^{2})^{m}}{m !}
\eeq
are Laguerre polynomials;
 $\xi$ is given  by
\beq
\xi = \sqrt{\frac{(M \Omega/2\hbar)^{2}}{1-\frac{M\omega_{c}}{2}\theta + \left(\frac{M \Omega}{4}\theta\right)^{2}}}
\eeq
and $n = 0, 1, 2, \dots$ is the principal quantum number while  $\rho = 0, \pm 1, \pm 2, \dots$ stands for the angular moment quantum number.

As already mentioned in the Introduction, we voluntarily choose to study the thermodynamics  of the system  faithfully following the analysis performed
in  \cite{gazeau-hsiao-jellal}. Although some main  expressions  appear similar in their form to those derived in that work, due to our system parameter reformulation,
this approach has  the interesting advantage to offer an easier relation comparison and to point up the contribution of the electric field not
considered in that work. On this basis,
from (\ref{eigens01}), we derive the thermodynamical potential using the formula:

\beq{\label{eigens02}}
\Gamma_{\theta}  = -\frac{1}{\beta}\sum_{n=0}^{\infty}\sum_{\rho=-\infty}^{\infty}
\log\left[1 + e^{-\beta(\mathcal E_{n,\rho, \theta} - \mu)}\right],
\eeq
where $\beta = 1/k_{B}T;$  $\mu$ is the chemical potential.

The resolution of the identity (\ref{resolv}) allows to apply the Berezin-Lieb inequalities \cite{feng-klauder-staryer-jpg,gazeau-hsiao-jellal}:

\beq
&&-\frac{1}{\beta \pi^{2}}\int_{\C^{2}}  \mbox{log}\left(1+ e^{-\beta(\mbox{\^{H}}_{q}- \mu)}\right)d^{2}z_{+}d^{2}z_{-}
\leq \Gamma_{\theta} \cr
&& \Gamma_{\theta} \leq
-\frac{1}{\beta \pi^{2}}\int_{\C^{2}}  \mbox{log}\left(1+ e^{-\beta(\mbox{\v{H}}_{q}- \mu)}\right)d^{2}z_{+}d^{2}z_{-}.
\eeq

By using the lower and  upper symbols of the Hamiltonian $H_{q},$  and
performing the angular integrations, where $u_{+} = |z_{+}|^{2}, v_{-} = |z_{-}|^{2}$ with
$z_{+} = re^{i\varphi}, \, z_{-} = \rho e^{i \phi}, \ r, \rho \geq 0,
\varphi, \phi \in [0,2\pi)$, we get

\beq{\label{ineq00}}
&& - \frac{1}{\beta} \int_{0}^{\infty}du_{+} \int_{0}^{\infty}dv_{-} \; \mbox{log}
(1+ e^{-\beta(\frac{\hbar}{2} \left(\tilde \Omega_{+}u_{+} + \tilde \Omega_{-}v_{-} - \tilde \Omega\right)
 - \mu_{e,E})}) \leq \Gamma_{\theta} \cr
&& \Gamma_{\theta} \leq
- \frac{1}{\beta} \int_{0}^{\infty}du_{+} \int_{0}^{\infty}dv_{-} \; \mbox{log}
(1+ e^{-\beta(\frac{\hbar}{2} \left(\tilde \Omega_{+}u_{+} + \tilde \Omega_{-}v_{-} + \tilde \Omega\right)
 -\mu_{e,E})})
\eeq

where $\mu_{e,E} = \mu + \frac{1}{2}e(E_{1} x_{0}  +
E_{2} y_{0})$.

Setting $u = \frac{\beta \hbar}{2}(\tilde \Omega_{+} u_{+} + \tilde \Omega_{-}v_{-}), \, v= \frac{\beta \hbar}{2}\tilde \Omega_{+} u_{+} $, performing
an integration by parts, and introducing the control parameters $\tilde \kappa'_{\pm}
= \mbox{exp}(\beta(\mu_{e,E} \pm \frac{\hbar \tilde \Omega}{2})) = \mbox{exp}(-\beta k) \cdot \tilde \kappa_{\pm}$, where
$\tilde \kappa_{\pm} = \mbox{exp}(\beta(\mu \pm \frac{\hbar \tilde \Omega}{2}))$, (\ref{ineq00}) is reduced to

\beq{\label{ineq01}}
\phi(\tilde \kappa'_{+}) \leq \Gamma_{\theta} \leq \phi(\tilde \kappa'_{-})
\eeq

where $\phi(\tilde \kappa')$ takes the form

\beq
\phi(\tilde \kappa') &=& -\frac{2 \tilde \kappa'}{\beta (\beta \hbar \omega_{0})^{2} }
\int_{0}^{\infty}\frac{u^{2} e^{-u}}{1 + \tilde \kappa' e^{-u}} du \cr
&=& \cases{
              \begin{array}{lll}
\frac{4 }{\beta (\beta \hbar \omega_{0})^{2} }F_{3}(-\tilde \kappa') \quad  \qquad  \qquad \qquad
\qquad \qquad \quad \mbox{for} \quad   \tilde \kappa' \leq 1 \\
\frac{4 }{\beta (\beta \hbar \omega_{0})^{2} } \left[ - \frac{(log \tilde \kappa' )^{3}}{6}  -
\frac{\pi^{2}log \tilde \kappa'}{6} + F_{3}(-\tilde \kappa'^{-1}) \right] \quad \mbox{for} \quad   \tilde \kappa' > 1
\end{array}}.
\eeq
The function $F_{s}$ is of the Riemann-Fermi-Dirac type \cite{gazeau-hsiao-jellal}:
\beq
F_{s}(z)= \sum_{n=1}^{\infty}\frac{z^{n}}{n^{s}}.
\eeq

In the high-temperature limit case, the assumption   $|\mu_{e,E} \pm \frac{\hbar \tilde \Omega}{2}| \gg \beta$  gives
$\tilde \kappa'_{\pm}  \approx 1$  so that using (\ref{ineq00}) and (\ref{ineq01}) the thermodynamical potential $\Gamma_{\theta}$
can be approximated by

\beq
\Gamma_{\theta} \approx \frac{4}{\beta^{3}\hbar^{2} }\frac{ F_{3}(-1)}{\omega^{2}_{0}} \approx -0,901543 \times
\frac{4}{\beta} \left(\frac{1}{\beta \hbar \omega_{0}}\right)^{2}.
\eeq

Consider the  expression of the function $\phi$ as follows \cite{gazeau-hsiao-jellal}:

\beq
\phi(\tilde \kappa'_{\pm}) = A \mp \frac{\Delta}{2} + S_{\pm}
\eeq

where in our considered physical model situation

\beq
A = -2\mu_{e,E} \left[\frac{1}{3}\left(\frac{\mu_{e,E}}{\hbar \omega_{0}}\right)^{2} +
\frac{1}{4}\left(\frac{\tilde \Omega}{\omega_{0}}\right)^{2} + \frac{\pi^{2}}{3}\left(\frac{1}{\beta \hbar \omega_{0}}\right)^{2}\right],
\eeq

\beq
\frac{\Delta}{2} = 2 \hbar \tilde \Omega \left[\frac{1}{2}\left(\frac{\mu_{e,E}}{\hbar \omega_{0}}\right)^{2} +
\frac{1}{24}\left(\frac{\tilde \Omega}{\omega_{0}}\right)^{2} + \frac{\pi^{2}}{6}\left(\frac{1}{\beta \hbar \omega_{0}}\right)^{2}\right],
\eeq

\beq
S_{\pm} = \frac{4 }{\beta (\beta \hbar \omega_{0})^{2} } F_{3}\left(-e^{-\beta(\mu_{e,E} \pm \frac{\hbar \tilde \Omega}{2})}\right).
\eeq

At low temperature, $S_{\pm}$ can be approximated by

\beq
S_{0} = \frac{4 }{\beta (\beta \hbar \omega_{0})^{2} } F_{3}\left(-e^{-\beta \mu_{e,E}}\right).
\eeq

Considering the following ratio

\beq
\frac{\Delta}{| A + S_{0}|} = \frac{\hbar \tilde \Omega}{\mu_{e,E}}\left[\frac{3 +
\pi^{2}\left(\frac{1}{\beta \mu_{e,E}}\right)^{2} + \frac{1}{4}\left(\frac{\hbar \tilde \Omega}{\mu_{e,E}}\right)^{2}}
{1 + \pi^{2}\left(\frac{1}{\beta \mu_{e,E}}\right)^{2} + \frac{3}{4}\left(\frac{\hbar \tilde \Omega}{\mu_{e,E}}\right)^{2}-
\left(\frac{1}{\beta \mu_{e,E}}\right)^{3} F_{3}\left(-e^{-\beta \mu_{e,E}}\right)}   \right]
\eeq

which tends to zero at low temperature, namely $\mu_{e,E} \gg \hbar \tilde \Omega / 2$ and $\mu_{e,E} \gg 1 / \beta$,
the thermodynamical potential can be obtained as

\beq
\Gamma_{\theta} & \approx & A + S_{0} \cr
&=& -2\mu_{e,E} \left[\frac{1}{3}\left(\frac{\mu_{e,E}}{\hbar \omega_{0}}\right)^{2}
+
\frac{1}{4}\left(\frac{\tilde \Omega}{\omega_{0}}\right)^{2} + \frac{\pi^{2}}{3}\left(\frac{1}{\beta \hbar \omega_{0}}\right)^{2} -
\frac{2}{\beta  \mu_{e,E} (\beta  \hbar \omega_{0})^{2} } F_{3}\left(-e^{-\beta \mu_{e,E}}\right)\right]. \nonumber \\
\cr
\eeq

The average number of electrons is given by

\beq
\langle N_{e} \rangle &\approx & -\partial_{\mu} (A + S_{0}) \cr
&= &
 4\left(\frac{\mu_{e,E}}{\hbar \omega_{0}}\right)^{2}\left[\frac{1}{2} + \frac{1}{8}\left(\frac{\hbar \tilde \Omega}{\mu_{e,E}}\right)^{2} +
\frac{\pi^{2}}{6}\left(\frac{1}{\beta \mu_{e,E}}\right)^{2} +
\left(\frac{1}{\beta \mu_{e,E}}\right)^{2}F_{2}\left(-e^{-\beta \mu_{e,E}}\right)\right] \cr
&\approx & 2\left(\frac{\mu_{e,E}}{\hbar \omega_{0}}\right)^{2} \quad  \mbox{for}  \quad
\mu_{e,E} \gg \hbar \tilde \Omega / 2 \quad  \mbox{and} \quad  \mu_{e,E} \gg 1 / \beta.
\eeq

The magnetic moment $\mathcal M_{\theta} = -\left(\frac{\partial \Gamma_{\theta}}{\partial B}\right)_{\mu}$ is derived as follows:

\beq{\label{magn00}}
\mathcal M_{\theta} &=& \frac{e \mu}{Mc}\left[\frac{\omega_{c}}{\omega^{2}_{0}} -
\frac{M \theta}{4\omega^{2}_{0}}(2\omega^{2}_{c} + \Omega^{2}) +
2 \frac{\omega_{c}}{\omega^{2}_{0}} \left(\frac{M  \Omega \theta}{4}\right)^{2}  \right] \cr
&& + \frac{e^{2}}{ 2Mc}(E_{1} x_{0}  +
E_{2} y_{0})
\left[\frac{\omega_{c}}{\omega^{2}_{0}} -
\frac{M \theta}{4\omega^{2}_{0}}(2\omega^{2}_{c} + \Omega^{2}) +
2 \frac{\omega_{c}}{\omega^{2}_{0}} \left(\frac{M  \Omega \theta}{4}\right)^{2}  \right]
\eeq

which provides the susceptibility  $\chi_{\theta} = \frac{\partial \mathcal M_{\theta}}{\partial B}$ obtained in the following form

\beq{\label{magn01}}
\chi_{\theta} &=& \left(\frac{e}{Mc\omega_{0}}\right)^{2} \mu\left[1- \frac{3}{2}M\theta \omega_{c} - (M\theta \omega_{0})^{2} +
 6\left(\frac{M  \Omega \theta}{4}\right)^{2} \right]  \cr
&& + \frac{e^{3}}{2(Mc\omega_{0})^{2}} (E_{1} x_{0}  +
E_{2} y_{0}) \left[1- \frac{3}{2}M\theta \omega_{c} - (M\theta \omega_{0})^{2} +
 6\left(\frac{M  \Omega \theta}{4}\right)^{2} \right].
\eeq

In both expressions  (\ref{magn00}) and (\ref{magn01}), the second terms stand for the contributions engendered by the presence of the electric field.

\begin{rmk}
In the absence of the electric field, this model is reduced to that described by the Fock-Darwin Hamiltonian  investigated in
the study
of the orbital magnetism of a two-dimensional noncommutative confined system \cite{jellal}. In that work,
 it has been shown that the degeneracy of Landau levels can be lifted via the $\theta$-term  at
weak magnetic field limit, i.e. for $\omega_{c} \ll \omega_{0}$.
\end{rmk}

By use of the Poisson summation formula \cite{landau3} in the sum over $n$ and $\rho$  in (\ref{eigens02}), we obtain

\beq
\Gamma_{\theta} = \Gamma^{0}_{\theta} + \Gamma^{L}_{\theta} + \Gamma^{OSC}_{\theta},
\eeq
where
\beq{\label{gamma0}}
\Gamma^{0}_{\theta} &=& -\frac{1}{\beta(\hbar \omega_{0})^{2}}
\int_{0}^{\infty} d\varepsilon \int_{0}^{\infty} d\eta
\log{(1 + e^{-\beta(\varepsilon + \eta
- \mu_{e, E})})} \crcr
&& + \frac{1}{12}\mu_{e,E},
\eeq

\beq{\label{gamma1}}
\Gamma^{L}_{\theta} = \frac{\mu_{e,E}}{24}
\left(\frac{\omega_{c} -\Theta_{M, \omega_{c}, \theta}}{\omega_{0}}\right)^{2},
\eeq

\beq{\label{gamma2}}
\Gamma^{OSC}_{\theta} &=&
\frac{1}{2 \pi \beta}\sum_{k=1}^{\infty}(-1)^{k} \left[\left(\frac{\tilde \Omega}{\omega_{0}}\right)^{2} \frac{1}{k^{2}}
- \frac{\pi^{2}}{3} \right] \times
\frac{\sin{\left\{2\pi_{\tilde \Omega}k\mu_{e,E}\right \}}}
{\mbox{Sinh}_{k,\tilde \Omega}} \crcr
&& + \frac{1}{2\pi \beta}\sum_{\sigma = \pm}\sum_{l=1}^{\infty}\frac{\tilde \Omega_{\sigma}}{\tilde \Omega}\frac{1}{l^{2}}
 \frac{\sin{\left\{2\pi_{\tilde \Omega_{\sigma}}l \mu_{e,E}\right \}}}
{\mbox{Sinh}_{l,\tilde \Omega_{\sigma}}} \crcr
&& + \frac{1}{\pi \beta} \sum_{\sigma = \pm} \sum_{k=1}^{\infty} \sum_{l=1}^{\infty}\frac{(-1)^{k}}{l} \crcr
&& \times [  \frac{\sin{\left\{\pi_{\tilde \Omega}\mu_{e, E}
K_{k,l,\tilde \Omega, \sigma;-}\right \}}
\cos{\left\{\pi_{\tilde \Omega}\mu_{e, E}
K_{k,l,\tilde\Omega, \sigma;+}\right \}}}
{K_{k,l,\tilde \Omega, \sigma;-}
\mbox{Sinh}_{l,\tilde \Omega_{\sigma}}} + \crcr
&&   \frac{\sin{\left\{\pi_{\tilde \Omega}\mu_{e, E}
K_{k,l,\tilde \Omega, \sigma;+}\right \}}
\cos{\left\{\pi_{\tilde \Omega}\mu_{e, E} K_{k,l,\tilde \Omega, \sigma;-}\right \}}}
{K_{k,l,\tilde \Omega, \sigma;+}\mbox{Sinh}_{l,\tilde \Omega_{\sigma}}}]  \nonumber \\
\eeq

with the relation $\mu \gg T$  assumed.

From the thermodynamical potential $\Gamma_{\theta}$, the magnetic moment
$\mathcal M_{\theta} = -\left(\frac{\partial \Gamma_{\theta}}{\partial B}\right)_{\mu}$ can be derived to give the following
contributions:

\beq
\mathcal M_{\theta}  =   \mathcal M^{L}_{\theta} + \mathcal M^{OSC}_{\theta,1} + \mathcal M^{OSC}_{\theta,2}
+ \mathcal M^{OSC}_{\theta,3},
\eeq

where
\begin{itemize}
\item
\beq
\mathcal M^{0}_{\theta}
&=& -\left(\frac{\partial \Gamma^{0}_{\theta}}{\partial B}\right)_{\mu} =0,
\eeq

\item
\beq
\mathcal M^{L}_{\theta} = -\left(\frac{\partial \Gamma^{L}_{\theta}}{\partial B}\right)_{\mu}
= -\frac{\mu_{e,E}}{12}\left[\frac{\omega_{c} -\Theta_{M, \omega_{c}, \theta}}{\omega^{2}_{0}} \right]\mathcal I_{e, B, \theta, M},
\eeq
\item
\beq
&&\mathcal M^{OSC}_{\theta,1} = -\frac{1}{2 \pi \beta} \sum_{k=1}^{\infty}(-1)^{k}[\quad \frac{1}{k^{2}\omega^{2}_{0}}
\left(\frac{2e\omega_{c}}{Mc}\left(\frac{\tilde \Omega}{\Omega}\right)^{2} +
\Omega^{2}B_{\theta}\right)\times \cr
&& \times \frac{\sin{\left\{2\pi_{\tilde \Omega}k\mu_{e,E}\right \}}}
{\mbox{Sinh}_{k,\tilde \Omega}} -
\frac{2\pi k}
{\hbar  \, \mbox{Sinh}_{k,\tilde \Omega}}
\left[\left(\frac{\tilde \Omega}{\omega_{0}}\right)^{2} \frac{1}{k^{2}}
 - \frac{\pi^{2}}{3}\right] \times \frac{\mathcal K_{\omega_{c}, e, \theta, B, M}}{\tilde \Omega^{2}}
 \times \cr
&& \times\left[\mu_{e, E} \cos{\left\{2\pi_{\tilde \Omega}k\mu_{e,E}\right \}} - \frac{\pi}{\beta}
\mbox{Coth}_{k, \tilde \Omega} \times \sin{\left\{2\pi_{\tilde \Omega}k\mu_{e,E}\right \}} \right]\quad ],
\eeq
\item
\beq
\mathcal M^{OSC}_{\theta,2} &=&  -\frac{1}{2 \pi \beta} \sum_{\sigma = \pm}\sum_{l=1}^{\infty}\frac{1}{l^{2}}
[\quad
[\pm
\frac{1}{\tilde \Omega^{2}}[\mathcal L_{e,B, \theta, M} -
\frac{\tilde \omega_{c}}{2}\mathcal K_{\omega_{c}, e, \theta, B, M}]]
 \times \frac{\sin\left\{2\pi_{\tilde \Omega_{\sigma}}l \mu_{e,E}\right \}}
{\mbox{Sinh}_{l,\tilde \Omega_{\sigma}}} \cr
&& - \frac{\tilde \Omega_{\sigma}}{\tilde \Omega}
\frac{2\pi l}
{\hbar \, \mbox{Sinh}_{l,\tilde \Omega_{\sigma}}} \times \frac{1}{2\tilde \Omega_{\sigma}^{2}}
\left[\mathcal K_{\omega_{c}, e, \theta, B, M} \pm
\mathcal I_{e,B, \theta, M} \right]\times  \cr
&&\times \left[\mu_{e, E} \cos{\left\{2\pi_{\tilde \Omega_{\sigma}}l \mu_{e,E}\right \}} - \frac{\pi}{\beta}
\mbox{Coth}_{l, \tilde \Omega_{\sigma}} \times
\sin{\left\{2\pi_{\tilde \Omega_{\sigma}}l \mu_{e,E}\right \}} \right]\quad ],
\eeq

\item

\beq
&&\mathcal M^{OSC}_{\theta,3} = - \frac{1}{\pi \beta} \sum_{\sigma = \pm} \sum_{k=1}^{\infty} \sum_{l=1}^{\infty}\frac{(-1)^{k}}{l}
 \crcr
&&\times[\quad [\frac{1}
{\left[K_{k,l,\tilde \Omega, \sigma;-}\mbox{Sinh}_{l,\tilde \Omega_{\sigma}}\right]}
\times \pi_{\tilde \Omega}\mu_{e, E}
[\left[k\cos\left\{2\pi_{\tilde \Omega}k\right \}-l \frac{\tilde \Omega}{\tilde \Omega_{\sigma}
} \cos{\left\{2\pi_{\tilde \Omega_{\sigma}}l \mu_{e,E} \right \}} \right] \times \cr
&& \times \left[-\frac{\mathcal K_{\omega_{c}, e, \theta, B, M}}{\tilde \Omega}
\right] \pm l \cos{\left\{2\pi_{\tilde \Omega_{\sigma}}l \mu_{e,E}\right \}}
\times \frac{1}{\tilde \Omega^{2}_{\sigma}}\left(\mathcal L_{\omega_{c}, e, \theta, B, M} - \frac{\tilde \omega_{c}}{2}
\mathcal K_{\omega_{c}, e, \theta, B, M}\right)] \;  + \cr
&& \; + \frac{1}
{\left[K_{k,l,\tilde \Omega, \sigma;+}\mbox{Sinh}_{l,\tilde \Omega_{\sigma}}\right]} \times
 \pi_{\tilde \Omega}\mu_{e, E} [\left[k\cos\left\{2\pi_{\tilde \Omega}k\right \}-l \frac{\tilde \Omega}{\tilde \Omega_{\sigma}
} \cos{\left\{2\pi_{\tilde \Omega_{\sigma}}l \mu_{e,E}\right \}} \right] \times \cr
&& \times \left[-\frac{\mathcal K_{\omega_{c}, e, \theta, B, M}}{\tilde \Omega}
\right] \mp l \cos{\left\{2\pi_{\tilde \Omega_{\sigma}}l \mu_{e,E}\right \}} \times \frac{1}{\tilde \Omega^{2}_{\sigma}}
\left(\mathcal L_{\omega_{c}, e, \theta, B, M} - \frac{\tilde \omega_{c}}{2}
\mathcal K_{\omega_{c}, e, \theta, B, M}\right)] \; - \cr
&&-\frac{\sin{\left\{\pi_{\tilde \Omega}\mu_{e, E}
K_{k,l,\tilde \Omega, \sigma;+}\right \}}\cos{\left\{\pi_{\tilde \Omega}\mu_{e, E}
K_{k,l,\tilde \Omega, \sigma;-}\right \}}}{\left[K_{k,l,\tilde \Omega, \sigma;+}\mbox{Sinh}_{l,\tilde \Omega_{\sigma}}
\right]^{2}} \times \cr
&&\times [\;\;  \mp
\frac{l}{(\tilde \Omega_{\sigma})^{2}}\left(\mathcal L_{e,B, \theta, M} -
\frac{\tilde \omega_{c}}{2}\mathcal K_{\omega_{c}, e, \theta, B, M} \right)\mbox{Sinh}_{l,\tilde \Omega_{\sigma}} -\cr
&& - K_{k,l,\tilde \Omega, \sigma;+}
\frac{2\pi^{2}l}
{\beta \hbar  }\mbox{Cosh}_{l, \tilde \Omega_{\sigma}} \times \frac{1}{2(\tilde \Omega_{\sigma})^{2}}
\left(\mathcal K_{\omega_{c}, e, \theta, B, M} \pm
\mathcal I_{e, B, \theta, M}\right) \; \; ] - \cr
&& - \frac{\sin{\left\{\pi_{\tilde \Omega}\mu_{e, E}
K_{k,l,\tilde \Omega, \sigma;-}\right \}}\cos{\left\{\pi_{\tilde \Omega}\mu_{e, E}
K_{k,l,\tilde \Omega, \sigma;+}\right \}}}{\left[K_{k,l,\tilde \Omega, \sigma;-}\mbox{Sinh}_{l,\tilde \Omega_{\sigma}}
\right]^{2}} \times \cr
&&\times [\; \; \pm
\frac{l}{(\tilde \Omega_{\sigma})^{2}}\left(\mathcal L_{e,B, \theta, M} -
\frac{\tilde \omega_{c}}{2}\mathcal K_{\omega_{c}, e, \theta, B, M}\right)\mbox{Sinh}_{l,\tilde \Omega_{\sigma}} - \cr
&& - K_{k,l,\tilde \Omega, \sigma;-}
\frac{2\pi^{2}l}
{\beta \hbar  } \mbox{Cosh}_{l, \tilde \Omega_{\sigma}} \times \frac{1}{2(\tilde \Omega_{\sigma})^{2}}
\left(\mathcal K_{\omega_{c}, e, \theta, B, M} \pm
\mathcal I_{e, B, \theta, M}\right)\; \; ] \quad ]
;
\eeq
\end{itemize}
with
\beq{\label{quant00}}
\Theta_{M, \omega_{c}, \theta} = \frac{M\omega^{2}_{c} \theta}{4} +  M \omega^{2}_{0}\theta,\qquad B_{\theta} = \frac{e\theta}{2c}\left(\frac{eB}{4c}\theta - 1\right),
\eeq

\beq{\label{quant02}}
\pi_{\tilde \Omega} = \frac{\pi}{\hbar \tilde \Omega}, \qquad \pi_{\tilde \Omega_{\sigma}}
=  \frac{\pi}{\hbar \tilde \Omega_{\sigma}},\qquad k \pm \frac{\tilde \Omega}{ \tilde \Omega_{\sigma}}l = K_{k,l,\tilde \Omega, \sigma;\pm},
\eeq

\beq{\label{quant04}}
\sinh\left\{\frac{2\pi^{2}l}
{\beta \hbar  \tilde \Omega_{\sigma}}\right\} =  \mbox{Sinh}_{l,\tilde \Omega_{\sigma}} , \qquad \sinh\left\{\frac{2\pi^{2}k}
{\beta \hbar  \tilde \Omega}\right\} = \mbox{Sinh}_{k,\tilde \Omega},
\eeq

\beq{\label{quant05}}
\cosh\left\{\frac{2\pi^{2}l}
{\beta \hbar  \tilde \Omega_{\sigma}}\right\} =  \mbox{Cosh}_{l,\tilde \Omega_{\sigma}} , \qquad \cosh\left\{\frac{2\pi^{2}k}
{\beta \hbar  \tilde \Omega}\right\} = \mbox{Cosh}_{k,\tilde \Omega},
\eeq

\beq{\label{quant06}}
\mathcal I_{e, B, \theta, M} = \frac{e}{Mc}\left(1 - \frac{eB}{2c}\theta\right),\qquad
\mathcal K_{\omega_{c}, e, \theta, B, M} = \frac{\omega_{c}
e\tilde \Omega}{Mc\Omega^{2}} + \frac{\Omega^{2}}{\tilde \Omega}\frac{e\theta}{4c}\left(\frac{eB}{4c}\theta - 1\right),
\eeq

\beq{\label{quant08}}
\mathcal L_{e,B, \theta, M} = \frac{e\tilde \Omega}{2Mc}\left(1 - \frac{eB}{2c}\theta\right).
\eeq

\begin{rmk}
In  the situation where the  $\theta-$ parameter  is switched off,  we obtain:

\beq
\Theta_{M, \omega_{c}, \theta = 0} = 0, \qquad B_{\theta = 0} = 0,
\eeq

\beq
\mathcal I_{e, B, \theta=0, M} = \frac{e}{Mc}, \qquad
\mathcal K_{\omega_{c}, e, \theta=0, B, M} = \frac{e\omega_{c}
}{Mc\Omega}, \qquad \mathcal L_{e,B, \theta= 0, M} = \frac{e \Omega}{2Mc},
\eeq
where  we get $\tilde \Omega \equiv \Omega, \quad \tilde \omega_{c} \equiv \omega_{c}$ and $\tilde \Omega_{\sigma} \equiv
 \Omega_{\sigma} = \frac{\Omega \pm \omega_{c}}{2}$. The term  $k_{e,E} = - \frac{1}{2}e(E_{1} x_{0}  +
E_{2} y_{0})$ giving  $\mu_{e,E}$
 contributes  to  the chemical potential  $\mu$.
The   expressions  (\ref{gamma0})-(\ref{gamma2}) are identified
 for $k_{e, E} = 0$ with
those derived in \cite{ishikawa-fukuyama} for the Hamiltonian describing a two-dimensional electrons confined by an isotropic
harmonic potential in a perpendicular magnetic field, linked to the  Landau problem in the
commutative case,  for which coherent states have been constructed on the  Fock
Hilbert space  \cite{gazeau-hsiao-jellal}. Besides, the analysis done in \cite{jan-scholtz} for an ideal fermion gas may also permit to
understand the  behavior of the studied physical system.

\end{rmk}
\section{Matrix vector coherent states for  $\mathbb H^{dim}$}
\subsection{Construction}

Consider the set of continuous mappings $F_{n}(\mathfrak Z): \mathcal M_{4}(\C)\rightarrow \mathcal B(\mathcal H_{c})$ satisfying

\beq{\label{bound00}}
0 < \mathcal N(\mathfrak Z) = \sum_{n \in \N} tr_{c} [|F_{n}(\mathfrak Z)|^{2}] < \infty,
\eeq
where $\mathcal M_{4}(\C)$ is the space of $4 \times 4$  complex  matrices.
Then, there follows  that the linear map given by
\beq
T(\mathfrak Z): \C^{4} &\rightarrow& \C^{4} \otimes \mathcal H_{c} \cr
\chi^{j} &\mapsto& T(\mathfrak Z)\chi^{j}  = (\mathcal N(\mathfrak Z))^{-1/2} \sum_{n \in \N} F_{n}(\mathfrak Z)
|\chi^{j}, n \rangle, \qquad j=1,2,3,4
\eeq
is bounded. Set
\beq{\label{bound02}}
F_{n}(\mathfrak Z)|\chi^{j}, \tilde n\rangle &=&
\frac{\mathfrak Z^{n} \bar{\mathfrak Z}^{\tilde n} }{\sqrt{R(n)R(\tilde n)}}|\chi^{j}, \tilde n\rangle
\eeq

where $\mathfrak Z = diag(z_{1}, z_{2}, z_{3}, z_{4})$,   $z_{j} = r_{j}e^{\imath \theta_{j}}$
with $r_{j} \geq 0, \theta_{j} \in [0,2\pi)$ and  $R(n) = n !\mathbb I_{4}$.  Let
$\mathfrak W = diag(w_{1}, w_{2}, w_{3}, w_{4}), w_{j} = \rho_{j}e^{\imath \varphi_{j}}$ where $\rho_{j} \geq 0, \varphi_{j} \in [0,2\pi)$ and set
$R(m) = m!\mathbb I_{4}$.
With this setup and by analogy with the constructions
provided in \cite{ali-englis-gazeau, thirulo,  ben-scholtz},
the set of vectors  formally given by

\beq{\label{ncvcs00}}
|\mathfrak Z, \mathfrak W, \tau,  j, \tilde n, \tilde m)
&=& (\mathcal N(\mathfrak Z, \mathfrak W))^{-1/2}\sum_{n,m=0}^{\infty}
\frac{\mathfrak Z^{n} \bar{\mathfrak Z}^{\tilde n} }{\sqrt{R(n)R(\tilde n)}}
\frac{\mathfrak W^{m} \bar{\mathfrak W}^{\tilde m} }{\sqrt{R(m)R(\tilde m)}} e^{-\imath \tau \tilde E_{n,m}}\cr
&&|\chi^{j}\rangle \otimes |\tilde n \rangle   \langle\tilde m| \otimes |m\rangle \langle n|
\eeq

forms a set of VCS on  $\C^{4} \otimes \mathcal H_{q} \otimes \mathcal H_{q}$. These states satisfy a normalization condition to unity given by

\beq{\label{normnvcs00}}
\sum_{j=1}^{4}\sum_{\tilde n, \tilde m = 0}^{\infty}
(\mathfrak Z, \mathfrak W, \tau,  j, \tilde n, \tilde m|\mathfrak Z, \mathfrak W, \tau,  j, \tilde n, \tilde m) = 1
\eeq

with

\beq
\mathcal N(\mathfrak Z, \mathfrak W) =
e^{2(r^{2}_{1} + \rho^{2}_{1})}  + e^{2(r^{2}_{2} + \rho^{2}_{2})} + e^{2(r^{2}_{3} + \rho^{2}_{3})} + e^{2(r^{2}_{4}+ \rho^{2}_{4})}.
\eeq

Let
$D = \{(z_{1}, z_{2},
z_{3}, z_{4})  \in \C^4  \,| \; \,  |z_{j}|< \infty, j=1,2,3,4 \}$,
$\mathcal D =
\{(w_{1}, w_{2},
w_{3}, w_{4})  \in \C^4  \,| \; \,  |w_{j}|< \infty, j=1,2,3,4 \}$.

\bpro
The VCS (\ref{ncvcs00})  satisfy on the quantum Hilbert space $\C^{4} \otimes \mathcal H_{q} \otimes \mathcal H_{q}$ a resolution of the
identity as follows:
\beq{\label{ncres03}}
&&\sum_{j=1}^{4}\sum_{\tilde m = 0}^{\infty}\sum_{\tilde n = 0}^{\infty}\frac{1}{\tilde m !\tilde n !} \cr
&&\int_{D \times \mathcal  D}
d\mu(\mathfrak Z, \mathfrak W)(\overrightarrow{\partial_{\bar z_{j}}})^{\tilde n}
(\overrightarrow{\partial_{\bar w_{j}}})^{\tilde m}[\mathcal N(\mathfrak Z, \mathfrak W)|\mathfrak Z, \mathfrak W, \tau,  j, \tilde n, \tilde m)
(\mathfrak Z, \mathfrak W, \tau,  j, \tilde n, \tilde m|]
(\overleftarrow{\partial_{z_{j}}})^{\tilde n}(\overleftarrow{\partial_{w_{j}}})^{\tilde m} \cr
&&
= \mathbb I_{4} \otimes \mathbb I_{q} \otimes \mathbb I_{q}
\eeq
where the measure $d\mu(\mathfrak Z, \mathfrak W)$ is given on  $D \times \mathcal D$  by

\beq{\label{ncres05}}
d\mu(\mathfrak Z, \mathfrak W) =   \frac{1}{(2\pi)^{8}}
\prod_{j=1}^{4}\lambda(r_{j}) \varpi(\rho_{j})dr_{j}d\rho_{j}
d\theta_{j}d\varphi_{j}.
\eeq
\epro

{\bf Proof:}

In order to prove (\ref{ncres03}), let us first expand the integrand as

\beq{\label{bound01}}
&&\sum_{j=1}^{4}(\overrightarrow{\partial_{\bar{z}_{j}}})^{\tilde n}
(\overrightarrow{\partial_{\bar{w}_{j}}})^{\tilde m}[\mathcal N(\mathfrak Z, \mathfrak W)|\mathfrak Z, \mathfrak W, \tau,  j, \tilde n, \tilde m)
(\mathfrak Z, \mathfrak W, \tau,  j, \tilde n, \tilde m|]
(\overleftarrow{\partial_{z_{j}}})^{\tilde n}(\overleftarrow{\partial_{w_{j}}})^{\tilde m} \cr
&=& \sum_{j=1}^{4}\sum_{n,n', m, m' =0}^{\infty}(\overrightarrow{\partial_{\bar{z}_{j}}})^{\tilde n}(\overrightarrow{\partial_{\bar{w}_{j}}})^{\tilde m}
(F_{n}(\mathfrak Z)F_{m}(\mathfrak W)|\chi^{j}\rangle \otimes |\tilde n \rangle   \langle\tilde m| \otimes |m\rangle \langle n| )\cr
&& \times (F_{n'}(\mathfrak Z)F_{m'}(\mathfrak W)|\chi^{j}\rangle \otimes |\tilde n\rangle  \langle\tilde m|
\otimes |m'\rangle \langle n'| )^{\dag}
(\overleftarrow{\partial_{z_{j}}})^{\tilde n}(\overleftarrow{\partial_{w_{j}}})^{\tilde m} \cr
&:=& \sum_{j=1}^{4}\sum_{n,n', m, m'=0 }^{\infty}[(\partial_{\bar{z}_{j}})^{\tilde n}(F_{n}(\mathfrak Z)(\partial_{\bar{w}_{j}})^{\tilde m}F_{m}(\mathfrak W)
|\chi^{j}, \tilde n, \tilde m) (\chi^{j}, \tilde n, \tilde m|\cr
&& \times (\partial_{z_{j}})^{\tilde n}((F_{n'}(\mathfrak Z))^{*}(\partial_{w_{j}})^{\tilde m}(F_{m'}(\mathfrak W))^{*}]\otimes
| m \rangle   \langle n|n'\rangle \langle m'|,
\eeq

call $\mathcal I$ the operator on the left hand side of (\ref{ncres03}) and
choose arbitrary vectors $\Psi, \Psi', \Phi, \Phi'$ on the Hilbert space $\mathcal H_{q} \otimes \mathcal H_{q}$. Then, using (\ref{bound01}),
we have

\beq
(\Psi, \Psi'|\mathcal I|\Phi,\Phi') &=&
\sum_{j=1}^{4}\sum_{\tilde m = 0}^{\infty}\sum_{\tilde n = 0}^{\infty}\frac{1}{\tilde m !\tilde n !} \int_{D \times \mathcal D}
d\mu(\mathfrak Z, \mathfrak W) \times  \cr
&& \times \sum_{n,n', m, m'=0}^{\infty} (\Psi|[(\partial_{\bar{z}_{j}})^{\tilde n}
(F_{n}(\mathfrak Z)(\partial_{\bar{w}_{j}})^{\tilde m}F_{m}(\mathfrak W)
|\chi^{j}, \tilde n, \tilde m) \cr
&& (\chi^{j}, \tilde n, \tilde m|
(\partial_{z_{j}})^{\tilde n}((F_{n'}(\mathfrak Z))^{*}(\partial_{w_{j}})^{\tilde m}(F_{m'}(\mathfrak W))^{*}]|\Phi) \otimes
(\Psi'| m \rangle   \langle n| n'\rangle \langle m'|\Phi').  \nonumber \\
\eeq

The use of the boundedness of the operator $T$ and of the fact that $\sum_{j=1}^{4}|\chi^{j}\rangle \langle \chi^{j}| = \mathbb I_{4}$ allows
 to interchange the sum over $j$ with the integral and the four sums over $n,n'$ and $m,m'$, respectively. Thus,

\beq{\label{ncres04}}
&& (\Psi, \Psi'|\mathcal I|\Phi,\Phi')
=\sum_{\tilde m = 0}^{\infty}\sum_{\tilde n = 0}^{\infty}\frac{1}{\tilde m !\tilde n !}\int_{D \times \mathcal D}
d\mu(\mathfrak Z, \mathfrak W)  \times \cr
&&\times \sum_{n,n', m, m'=0}^{\infty} [\sum_{j=1}^{4}
(\Psi|\chi^{j}\rangle[(\partial_{\bar{z}_{j}})^{\tilde n}(F_{n}(\mathfrak Z)(\partial_{\bar{w}_{j}})^{\tilde m}F_{m}(\mathfrak W)
|\tilde n, \tilde m)
 \cr
&& (\tilde n, \tilde m|
(\partial_{z_{j}})^{\tilde n}((F_{n'}(\mathfrak Z))^{*}(\partial_{w_{j}})^{\tilde m}(F_{m'}(\mathfrak W))^{*}]\langle \chi^{j}|\Phi)]
\otimes (\Psi'| m \rangle   \langle n|n'\rangle \langle m'|\Phi') \cr
&=& \sum_{\tilde m = 0}^{\infty}\sum_{\tilde n = 0}^{\infty}\frac{1}{\tilde m !\tilde n !} \int_{D \times \mathcal  D}
d\mu(\mathfrak Z, \mathfrak W)
\sum_{n,n', m, m'=0}^{\infty}\cr
&&diag(\frac{ \tilde n! z^{n}_{1}}{\sqrt{n! \tilde n!}}
\frac{\tilde m!w^{m}_{1}}{\sqrt{m! \tilde m!}}, \dots,
\frac{\tilde n!z^{n}_{4}}{\sqrt{n! \tilde n!}}
\frac{\tilde m! w^{m}_{4}}{\sqrt{m! \tilde m!}}) \times diag(\frac{\tilde n!\bar z^{n'}_{1}}{\sqrt{n'! \tilde n!}}
\frac{\tilde m!\bar w^{m'}_{1}}{\sqrt{m'! \tilde m!}}, \dots, \frac{\tilde n!\bar z^{n'}_{4}}{\sqrt{n'! \tilde n!}}
\frac{\tilde m!\bar w^{m'}_{4}}{\sqrt{m'! \tilde m!}})\cr
&&  (\Psi|\tilde n, \tilde m) (\tilde n, \tilde m|\Phi)\otimes (\Psi'|n',m') (n,m|\Phi') \cr
&=& \sum_{m,\tilde m = 0}^{\infty}\sum_{n,\tilde n = 0}^{\infty} (\Psi|diag(2\pi \int_{0}^{\infty}r_{1}dr_{1}W(r_{1})\frac{r^{2 n}_{1}}{ n !}
2\pi \int_{0}^{\infty}\rho_{1}d\rho_{1}\tilde W(\rho_{1})\frac{\rho^{2 m}_{1}}{ m !},\cr
&& \dots, 2\pi \int_{0}^{\infty}r_{4}dr_{4}W(r_{4})\frac{r^{2 n}_{4}}{n !}
2\pi \int_{0}^{\infty}\rho_{4}d\rho_{4}\tilde W(\rho_{4})\frac{\rho^{2 m}_{4}}{ m !})
\langle \tilde n|\tilde n\rangle \langle\tilde m|\tilde m \rangle |\Phi) \cr
&&\otimes
(\Psi'|\langle n |n\rangle  \langle m|m\rangle |\Phi') \cr
\cr
&=& (\Psi,\Psi'|\Phi,\Phi')
\eeq
where the  moment problems  are solved by
$W(r_{j}) = (1/2\pi) \lambda(r_{j}), \tilde W(\rho_{j}) =
(1/2\pi)\varpi(\rho_{j})$ with $\lambda(r_{j}) = 2e^{-r^{2}_{j}}, \, \varpi(\rho_{j}) = 2e^{-\rho^{2}_{j}}$, respectively.
$\hfill{\square}$

There also results this:

\bpro
These states  fulfill the following properties:

\bitem
\item [i-]Temporal stability
\beq
\mathbb U(t)|\mathfrak Z, \mathfrak W, \tau,  j, \tilde n, \tilde m)  = |\mathfrak Z, \mathfrak W, \tau + t,  j, \tilde n, \tilde m),
\qquad \mathbb U(t) = e^{-\imath t \mathbb H^{dim}}.
\eeq
\item [ii-]Action identity
\beq
\sum_{j=1}^{4}\sum_{\tilde n, \tilde m = 0}^{\infty}
(\mathfrak Z, \mathfrak W, \tau,  j, \tilde n, \tilde m|\mathbb H^{dim}|\mathfrak Z, \mathfrak W, \tau,  j , \tilde n, \tilde m)  =
\frac{1}{2}\left(\frac{\tilde \Omega_{+}}{\tilde \Omega}|\mathfrak Z|^{2} + \frac{\tilde \Omega_{-}}{\tilde \Omega}
|\mathfrak W|^{2}+ 1\right).
\eeq

\eitem
\epro

\subsection{Quaternionic vector coherent states}

\subsubsection{Construction}
We briefly discuss now the QVCS construction and their  connection with  the studied  VCS.
In (\ref{ncvcs00}),
set $\mathfrak Z = diag(z,\bar z, z, \bar z)$ and $\mathfrak W = diag(w, \bar w, w, \bar w)$ where
$z = re^{-\imath \tilde \phi}, w = \rho e^{-\imath \tilde \varphi}$ with $r, \rho \geq 0, \, \tilde \phi, \tilde \varphi \in [0, 2\pi)$.
Consider  $u, v \in SU(2)$ and take $\mathcal Z = U \mathfrak Z U^{\dag}, \mathcal W = V \mathfrak W V^{\dag}$ where
$U = diag(u,u), \, V = diag(v,v)$.  Introduce the quaternions
$\mathfrak q = A(r)e^{\imath \vartheta \Theta(\hat n )}, \mathfrak Q = B(\rho)e^{\imath \gamma \tilde \Theta(\hat k) }$ with
$\Theta(\hat n ) = diag(\sigma(\hat n ), \sigma(\hat n )) , \, 
\tilde\Theta(\hat k ) = diag(\tilde \sigma(\hat k ), \tilde \sigma(\hat k ))$,  where $A(r) = r\mathbb I_{4},
\, B(\rho) = \rho\mathbb I_{4}$ and
\beq
\sigma(\hat n ) = \left(\begin{array}{cc}
\cos{\phi} & e^{\imath \eta}\sin{\phi} \\
e^{-\imath \eta}\sin{\phi} & -\cos{\phi}
                        \end{array}
\right), \quad \tilde \sigma(\hat k) = \left(\begin{array}{cc}
\cos{\varphi} & e^{\imath \varrho}\sin{\varphi} \\
e^{-\imath \varrho}\sin{\varphi}  & -\cos{\varphi}
                                             \end{array}
\right)
\eeq
where $\phi, \varphi \in [0, \pi]$ and $\vartheta, \gamma, \eta,\varrho \in [0, 2\pi)$.

From the scheme developed in  \cite{thirulogasanthar-ali}, since $u, v$ are given as $u = u_{\xi_{1}}u_{\phi_{1}}u_{\xi_{2}}, \,
v = v_{\zeta_{1}}v_{\phi_{2}}v_{\zeta_{2}}$ with $ u_{\xi_{1}} = diag(e^{\imath \xi_{1}/2}, e^{-\imath \xi_{1}/2}), \,
u_{\xi_{2}} = diag(e^{\imath \xi_{2}/2}, e^{-\imath \xi_{2}/2}), \,  v_{\zeta_{1}} = diag(e^{\imath \zeta_{1}/2}, e^{-\imath \zeta_{1}/2}),
 v_{\zeta_{2}} = diag(e^{\imath \zeta_{2}/2}, e^{-\imath \zeta_{2}/2})$, and
\beq
u_{\phi_{1}} = \left(\begin{array}{cc}
\cos{\frac{\phi_{1}}{2}} & \imath \sin{\frac{\phi_{1}}{2}} \\
\imath \sin{\frac{\phi_{1}}{2}} & \cos{\frac{\phi_{1}}{2}}
                     \end{array}
\right), \qquad v_{\phi_{2}} = \left(\begin{array}{cc}
\cos{\frac{\phi_{2}}{2}} & \imath \sin{\frac{\phi_{2}}{2}} \\
\imath \sin{\frac{\phi_{2}}{2}} & \cos{\frac{\phi_{2}}{2}}
                     \end{array}
\right), \quad \xi_{1}, \xi_{2}, \zeta_{1}, \zeta_{2} \in [0,2\pi),
\eeq

for $ \xi_{1}= \xi_{2} = \eta$ and $\zeta_{1}=  \zeta_{2} = \varrho$, we get the following identifications:
$\mathcal Z = r(\mathbb I_{4}\cos{\vartheta} + \imath
\Theta(\hat n )\sin{\vartheta}) = \mathfrak q, \,
\mathcal W = \rho(\mathbb I_{4}\cos{\gamma} + \imath \tilde\Theta(\hat k )\sin{\gamma}) = \mathfrak Q$.

 Then, the QVCS obtained as $|U\mathfrak Z U^{\dag}, V \mathfrak  W V^{\dag}, \tau, j, \tilde n, \tilde m) =
|\mathfrak q, \mathfrak Q,  \tau, j, \tilde n, \tilde m)$ can be written as
\beq{\label{nqvcs00}}
|\mathfrak q, \mathfrak Q,  \tau, j, \tilde n, \tilde m) =
(\mathcal N(r, \rho))^{-1/2}\sum_{n,m=0}^{\infty}
\frac{\mathfrak q^{n} \bar{\mathfrak q}^{\tilde n} }{\sqrt{n !\tilde n !}}
\frac{\mathfrak Q^{m} \bar{\mathfrak Q}^{\tilde m} }{\sqrt{m !\tilde m !}} e^{-\imath \tau \tilde E_{n,m}}
|\chi^{j}\rangle \otimes | \tilde n \rangle   \langle \tilde m| \otimes |m\rangle \langle n|.
\eeq

They satisfy a normalization  condition to unity given by

\beq
\sum_{j=1}^{4}\sum_{\tilde n, \tilde m = 0}^{\infty}
(\mathfrak q, \mathfrak Q, \tau,  j, \tilde n, \tilde m|\mathfrak q, \mathfrak Q, \tau,  j, \tilde n, \tilde m) = 1
\eeq

which provides $\mathcal N(r, \rho) =  4e^{2(r^{2} + \rho^{2})}$.
\bpro
The QVCS (\ref{nqvcs00}) fulfill a resolution of the identity property on $\C^{4} \otimes \mathcal  H_{q} \otimes \mathcal  H_{q}$ given by
\beq{\label{qvcsresolu}}
&&\sum_{j=1}^{4}\sum_{\tilde m = 0}^{\infty}\sum_{\tilde n = 0}^{\infty}\frac{1}{\tilde m !\tilde n !} \times \cr
&&\times \int_{D_{1} \times D_{2}}
d\mu(\mathfrak q, \mathfrak Q)(\overrightarrow{\partial_{r}})^{\tilde n}
(\overrightarrow{\partial_{\rho}})^{\tilde m}[W(r,\rho)|\mathfrak q, \mathfrak Q, \tau,  j, \tilde n, \tilde m)
(\mathfrak q, \mathfrak Q, \tau,  j, \tilde n, \tilde m|]
(\overleftarrow{\partial_{r}})^{\tilde n}(\overleftarrow{\partial_{\rho}})^{\tilde m} \cr
&&
= \mathbb I_{4} \otimes \mathbb I_{q} \otimes \mathbb I_{q}
\eeq

where $d\mu(\mathfrak q, \mathfrak Q) = \frac{1}{16 \pi^{2}}rdr \rho d\rho (\sin{\phi})d\phi  d\eta d\vartheta (\sin{\varphi})d\varphi
 d\varrho d\gamma$ on $D_{1} \times D_{2};$
\epro
$D_{1} = \{(r,\phi, \eta, \vartheta)| 0 \leq r < \infty,  0 \leq \phi \leq \pi,  0 \leq \eta, \vartheta < 2\pi\}$ and
$D_{2} = \{(\rho,\varphi, \varrho, \gamma)| 0 \leq \rho < \infty,  0 \leq \varphi \leq \pi,  0 \leq \varrho, \gamma < 2\pi\}$.

 The moment problem issued from (\ref{qvcsresolu}), stated as follows:
\beq
\int_{0}^{\infty} \int_{0}^{\infty} \frac{ 4 \pi^{2} W(r, \rho)}{ \mathcal N(r, \rho)}
\frac{r^{2 n}}{ n !}
\frac{\rho^{2 m}}{ m !} rdr  \rho d\rho =  1
\eeq
is solved
with

\beq
 W(r, \rho) = \frac{1}{\pi^{2}} \mathcal N(r, \rho)e^{-(r^{2}+\rho^{2})}.
\eeq

A connection with the Weyl-Heisenberg group is realized by considering the unitary operators  given by
\beq
U_{R}(0, \mathfrak q) &=& e^{\left[-\mathfrak q \otimes a^{\dag}_{R}  + \mathfrak q^{\dag} \otimes a_{R}\right]}  =
e^{-1/2\left[-\mathfrak q \otimes a^{\dag}_{R},    \mathfrak q^{\dag} \otimes a_{R}\right]}
e^{\mathfrak q^{\dag} \otimes a_{R}} e^{- \mathfrak q \otimes a^{\dag}_{R}} \cr
U_{L}(0, \mathfrak Q) &=& e^{\left[\mathfrak Q \otimes d^{\dag}_{L}  - \mathfrak Q^{\dag} \otimes d_{L}\right]}  =
e^{-1/2\left[\mathfrak Q \otimes d^{\dag}_{L},    -\mathfrak Q^{\dag} \otimes d_{L}\right]}
 e^{\mathfrak Q \otimes d^{\dag}_{L}} e^{-\mathfrak Q^{\dag} \otimes d_{L}}
\eeq

such that

\beq
|\mathfrak q, \mathfrak Q, \tau, j, \tilde n, \tilde m) =
\frac{e^{-(r^{2} + \rho^{2})/2}}{2}\mathbb U(\tau)\left[
\frac{\bar{\mathfrak q}^{\tilde n} \bar{\mathfrak Q}^{\tilde m} }{\sqrt{\tilde n !\tilde m !}}
|\chi^{j}\rangle \otimes |\tilde n\rangle  \langle\tilde m|
\otimes  U_{L}(0, \mathfrak Q)|0 \rangle \langle 0|U_{R}(0, \mathfrak q)\right],
\eeq

where the operators $a_{R}$ and $d_{L}$ act on a given   state $| \tilde n \rangle   \langle \tilde m| \otimes
|m\rangle \langle n|$  as follows:

\beq{\label{vec}}
a_{R}| \tilde n \rangle   \langle \tilde m| \otimes
|m\rangle \langle n|  &:=& | \tilde n \rangle   \langle \tilde m| \otimes
|m\rangle \langle n|a \cr
&=& \sqrt{n+1}|\tilde n \rangle  \langle \tilde m| \otimes |m \rangle \langle  n+1|,  \cr
 d_{L}| \tilde n \rangle   \langle \tilde m| \otimes
|m\rangle \langle n|  &:=& | \tilde n \rangle   \langle \tilde m| \otimes
d|m\rangle \langle n| \cr
&=& \sqrt{m}|\tilde n \rangle  \langle \tilde m| \otimes |m-1 \rangle \langle  n|.
\eeq

\subsubsection{QVCS statistical properties}

Let us consider the  operators given  on $\C^{4} \otimes \mathcal  H_{q} \otimes \mathcal  H_{q}$ by

\beq
\hat P_{X} = \mathbb I_{4} \otimes \frac{-\imath \hbar }{\sqrt{2 \theta}}[a_{R} - a^{\dag}_{R}, \ .],
\qquad
\hat P_{Y} =  \mathbb I_{4} \otimes \frac{-\hbar }{\sqrt{2 \theta}}[a_{R} + a^{\dag}_{R}, \ .],
\eeq

\beq
\hat X  =  \mathbb I_{4} \otimes \sqrt{\frac{\theta}{2}}[a_{R}+a^{\dag}_{R}], \qquad
\hat Y = \mathbb I_{4} \otimes \imath \sqrt{\frac{\theta}{2}}[a^{\dag}_{R} - a_{R}].
\eeq

From (\ref{vec}), we obtain

\beq
[a_{R} - a^{\dag}_{R}, \ |\tilde n \rangle  \langle \tilde m| \otimes | m \rangle\langle n|] =
\sqrt{n+1}|\tilde n \rangle  \langle \tilde m| \otimes |m \rangle \langle n+1| -
\sqrt{n}|\tilde n \rangle  \langle \tilde m| \otimes  |m \rangle \langle n|.
\eeq

Denote the expectation value
of an operator by
$\langle \cdot \rangle = \sum_{\tilde n, \tilde m  = 0}^{\infty}
(\mathfrak q, \mathfrak Q, j,  \tilde n,  \tilde m|\cdot|\mathfrak q, \mathfrak Q, j,  \tilde n,  \tilde m)$. We get
the following expressions:

\beq{\label{quad00}}
\langle \hat P_{X} \rangle
 = \pm\frac{\hbar }{2\sqrt{2 \theta}}r\cos{(\phi) \sin{(\eta)}}, \qquad \langle \hat P^{2}_{X} \rangle =
\frac{\hbar^{2} }{2 \theta}[r^{2}\sin^{2}(\eta) +\frac{1}{4}],
\eeq

\beq{\label{quad02}}
\langle \hat P_{Y} \rangle =  -
\frac{\hbar }{2\sqrt{2 \theta}}[r \cos (\eta) ], \qquad \langle \hat P^{2}_{Y} \rangle =
\frac{\hbar^{2} }{2 \theta}[r^{2}\cos^{2}(\eta) + \frac{1}{4}]
\eeq
from which result the relations

\beq{\label{quad03}}
(\Delta \hat P_{X} )^{2}
&=& \frac{1}{4}\left(\frac{\hbar^{2} }{2 \theta}\right) [4r^{2}\sin^{2}(\eta) - r^{2}\cos^{2}(\phi)\sin^{2}(\eta)  +1], \cr
(\Delta \hat P_{Y} )^{2}
&=& \frac{1}{4}\left(\frac{\hbar^{2} }{2 \theta}\right) [3r^{2}\cos^{2}(\eta) +1].
\eeq

In the same way, we obtain

\beq
(\Delta \hat Y)^{2}
&=& \frac{1}{4}\left(\frac{\theta}{2}\right) [4r^{2}\sin^{2}(\eta) - r^{2}\cos^{2}(\phi)\sin^{2}(\eta)  +1], \cr
(\Delta \hat X)^{2}
&=& \frac{1}{4}\left(\frac{\theta}{2}\right) [3r^{2}\cos^{2}(\eta) +1]
\eeq
and  the following uncertainties

\beq
[\Delta \hat X \Delta \hat Y]^{2} &=&  \frac{1}{16}\left(\frac{\theta^{2}}{4}\right)   F(r,\eta,\phi) =
\frac{1}{16} \left[\frac{1}{4}|\langle [\hat X, \hat Y] \rangle|^{2} \right]F(r,\eta,\phi),  \cr
[\Delta \hat X  \Delta \hat P_{X}]^{2}  &=&    \frac{1}{16} \left(\frac{\hbar^{2} }{4}\right)
 F(r,\eta,\phi) \geq  \frac{1}{16} \left[  \frac{1}{4}|\langle [\hat X, \hat P_{X}]  \rangle|^{2}\right],  \cr
[\Delta \hat Y \Delta \hat P_{Y}]^{2}  & = &   \frac{1}{16} \left(\frac{\hbar^{2} }{4}\right)
 F(r,\eta,\phi) \geq  \frac{1}{16} \left[  \frac{1}{4}|\langle [\hat Y, \hat P_{Y}]  \rangle|^{2}\right], \cr
[\Delta \hat P_{X} \Delta \hat P_{Y}]^{2} &= & \frac{1}{16} \left(\frac{\hbar^{4} }{4\theta^{2}}\right)
F(r,\eta,\phi) \geq \frac{1}{16}
\left[  \frac{1}{4}|\langle [\hat P_{X}, \hat P_{Y}]  \rangle|^{2}\right] = 0,
\eeq
where
\beq
F(r,\eta,\phi) = [3r^{2}\cos^{2}(\eta) +1]
[4r^{2}\sin^{2}(\eta) - r^{2}\cos^{2}(\phi)\sin^{2}(\eta)  +1].
\eeq

\begin{rmk} As a matter of result checking, let us attract the reader attention on the fact that,
in \cite{ben-scholtz},  a factor
  \textquotedblleft 2 \textquotedblright  has been forgotten
in the author expressions (\ref{quad00})-(\ref{quad03}) obtained for the dispersions of the momentum operators; one should read, in the denominator,
  the quantity
$\hbar^{2}/2\theta$ instead of $\hbar^{2}/\theta.$
Furthermore, in the mentioned work,  a sign \textquotedblleft - \textquotedblright should be also added in the expression of the operator
$\hat P_{Y}.$
\end{rmk}

\section{Concluding remarks}
A matrix formulation of a Hamiltonian describing the motion of an electron in an electromagnetic field with a confining harmonic potential
in a two-dimensional noncommutative space has been provided in this work. Relevant thermodynamical and statistical properties of the physical system
have been studied and
 discussed. In this analysis, some $\theta-$ modified quantities have been obtained. In the limit $\theta \rightarrow 0$, these
quantities can be identified with those derived in the commutative context related to the standard Landau problem.
Then, the MVCS have been constructed and  analyzed with respect to required properties.
Finally, the QVCS as well as their connection with the VCS and their statistical properties have been investigated and discussed.

\section*{Acknowledgements}
This work is partially supported by the Abdus Salam International
Centre for Theoretical Physics (ICTP, Trieste, Italy) through the
Office of External Activities (OEA) - \mbox{Prj-15}. The ICMPA
is in partnership with
the Daniel Iagolnitzer Foundation (DIF), France.


\begin{thebibliography}{10}
\addcontentsline{toc}{chapter}{References}
\bibitem{doplicher}
Doplicher S, Fredenhagen K and Roberts J E, {\it Commun. Math. Phys.} {\bf 172}  187 (1995)
\bibitem{connes-schwartz}
Connes A, Douglas M R and  Schwarz A, {\it JHEP} {\bf 9802}  003  (1998)
\bibitem{seiberg-witten}
Seiberg N and Witten E, {\it JHEP} {\bf 9909}  032  (1999)

\bibitem{douglas}
Douglas M R and  Nekrasov N A, {\it Rev. Mod. Phys.} {\bf 73}  97  (2001)
\bibitem{omer-jellal1}
Dayi  \"{O} F and Jellal A, {\it Phys. Lett.} A  {\bf 287}  349  (2001)
\bibitem{horvathy-duval}
Duval C and Horv\'{a}thy P A, {\it The exotic Galilei group and the \textquotedblleft Peierls substitution \textquotedblright}, 
 {\it Phys. Lett.}  B {\bf 479} 284 (2000)
\bibitem{horvathy}
Horv\'{a}thy P A, {\it The noncommutative Landau problem}, {\it Ann. Phys.} {\bf 299}  128 (2002)
\bibitem{omer-jellal}
Dayi  \"{O} F and Jellal A,
\emph{Hall effect in noncommutative coordinates},
{\it J. Math. Phys.} {\bf 43}  4592 (2002)
\bibitem{pasquier}
Pasquier V,
\emph{Quantum Hall Effect and Noncommutative Geometry}, S\'{e}minaire Poincar\'{e} X (2007)
\bibitem{nair-polychronakos}
Nair V P and Polychronakos A P,   {\it Phys. Lett.}  B {\bf 505} 267 (2001)
\bibitem{goerbig}
Goerbig M O, Lederer P and Smith C M,   \emph{Solides et liquides quantiques, dans les
syst\`{e}mes bidimensionnels d'\'{e}lectrons}, Quanta et photons, 105-110 (2005)
\bibitem{lederer}
Goerbig M O and  Lederer P,  \emph{Introduction to the quantum Hall effects}, Lecture notes (2006)
\bibitem{landau}
Landau L D,     {\it Z. Phys.} {\bf 64}  629 (1930)
\bibitem{jellal}
Jellal A, {\it J. Phys. A: Math. Gen.} {\bf 34} 10159  (2001)
\bibitem{gamboa-loewe-mendez-rojas}
Gamboa J, Loewe M, M\'{e}ndez F and  Rojas J C, \emph{The Landau problem in  noncommutative 
 Quantum Mechanics}, {\it Mod. Phys. Lett. A} {\bf 16}, 2075 (2001), e-preprint
arXiv:hep-th/0104224 (2001)
\bibitem{geloun-jan-hounkonnou}
Ben Geloun J, Govaerts J and Hounkonnou M N, \emph{A (p, q)-deformed Landau problem in a spherical harmonic well: spectrum and noncommuting 
coordinates},  {\it EPL} {\bf 80} 30001 (2007)
\bibitem{dulat-li}
Dulat S and Li K, \emph{Landau Problem in Noncommutative Quantum Mechanics}, {\it Chin. Phys.} C {\bf 32} 92 (2008)

\bibitem{alvarez}
Alvarez P D, Gomis J, Kamimura K and Plyushchay M S, \emph{Anisotropic harmonic oscillator, non-commutative Landau problem and exotic Newton-Hooke 
Symmetry},  {\it Phys. Lett.} B {\bf 659} 906 (2008) [arXiv: 0711.2644]
\bibitem{ben-sunandan-scholtz}
Ben Geloun J, Gangopadhyay S and Scholtz F G,
\emph{Harmonic oscillator in a background magnetic field in noncommutative quantum phase-space}, {\it EPL} {\bf 86} 51001 (2009)
\bibitem{zhang-horvathy}
Zhang P-M and Horv\'{a}thy P A, \emph{Chiral decomposition in the non-commutative Landau problem}, arXiv: hep-th/1112.0409,   {\it Ann. Phys.},  
{\it in press} (2012)
\bibitem{scholtz-chakraborty-jan-vaidya}
Scholtz F G, Chakraborty B, Govaerts J and Vaidya S,   {\it J. Phys. A: Math. Theor.}  {\bf 40}  14581 (2007)
\bibitem{jan-scholtz}
Scholtz F G and  Govaerts J,
\emph{Thermodynamics of a non-commutative fermion gas}, {\it J. Phys. A: Math. Theor.} {\bf 41} 505003 (2008)
\bibitem{gazeau-hsiao-jellal}
Gazeau J P,  Hsiao P Y and Jellal A, \emph{A Coherent-State Approach to Two-dimensional Electron Magnetism}, {\it Phys. Rev. B} {\bf 65} 094427 (2002)
\bibitem{ishikawa-fukuyama}
Ishikawa Y and Fukuyama H, {\it J. Phys. Soc. Jpn.} {\bf 68}  2405 (1999)
\bibitem{gouba-scholtz}
Scholtz F G, Gouba L, Hafver A and Rohwer C M,
{\it J. Phys. A: Math. Theor.} {\bf 42}  175303 (2009)
\bibitem{ben-scholtz}
Ben Geloun J and  Scholtz F G,
\emph{Coherent states in noncommutative quantum mechanics},
{\it J. Math. Phys.} {\bf 50} 043505 (2009)
\bibitem{ali-englis-gazeau}
Ali S T, Engli$\rm\check{s}$ M and  Gazeau J P,  {\it J. Phys. A: Math. Gen.} {\bf 37} 6067 (2004)
\bibitem{thirulo}
Thirulogasanthar K, Honnouvo G and Krzy\.{z}ak A, {\it Annals of Physics} {\bf 314} 119 (2004)
\bibitem{feng-klauder-staryer-jpg}
Feng D H, Klauder J R  and  Strayer M (eds.), {\it Coherent States: Past, Present and
Future} (Proc. Oak Ridge 1993),  World Scientific, Singapore (1994)
\bibitem{landau3}
Landau L  D  and  Lifshitz E  M, \emph{Statistical Physics, Part $1$} (Oxford:Pergamon) (1980)
\bibitem{thirulogasanthar-ali}
Ali S T and  Thirulogasanthar K,  {\it J. Math. Phys.} {\bf 44}  5070 (2003)
\end{thebibliography}
\end{document}